\documentclass[12p]{article}
 \usepackage{dcolumn}

\usepackage{latexsym}
\usepackage{cite}
\usepackage{color}
\usepackage{amsmath}
\usepackage{epsfig}
\usepackage{hyperref}

\newcommand{\ortala}[1]{\begin{center}#1\end{center}}

\newcommand{\sandd}[1]{\left\langle #1\right\rangle}
\newcommand{\sanddr}[1]{\left\langle\left\langle #1\right\rangle\right\rangle_r}

\newcommand{\summ}[3]{{{\underset{#1 }{\overset{#2}{\displaystyle\sum}}}#3}}

\newcommand{\re}[1]{(\ref{#1})}

\newcommand{\eq}[2]{\begin{equation}\label{#1}  #2\end{equation}}

\newcommand{\paran}[1]{\left(#1\right)}

\newcommand{\sch}[1]{Schrodinger}
\newcommand{\komb}[2]{\paran{\begin{array}{c} #1 \\ #2 \end{array}}}

\newcommand{\sanddrtek}[1]{\left\langle\left\langle 
#1\right\rangle\right\rangle_{r}}
\linethickness{0.6pt}

\begin{document}

\ortala{\textbf{Hysteresis Characteristics of Generalized Spin-S Magnetic Binary Alloys}}

\ortala{G\"ul\c{s}en  Karakoyun\footnote{gulsennkarakoyun@gmail.com}}
\ortala{\textit{The Graduate School of Natural and Applied Sciences, Dokuz 
Eyl{\"u}l University, Tr-35160 {\.I}zmir, Turkey}}

\ortala{\"Umit Ak\i nc\i\footnote{umit.akinci@deu.edu.tr}}
\ortala{\textit{Department of Physics, Dokuz Eyl\"ul University,
TR-35160 Izmir, Turkey}}

\section{Abstract}\label{Abstract}


In this study, hysteresis characteristics of the generalized spin-S binary alloy represented by the 
formula $A_c B_{1-c}$ have been investigated within the framework of effective field approximation. 
The binary system consists of type A (spin-S) and type B (spin-S) atoms which are randomly distributed 
on a honeycomb lattice. Both integer and half-integer spin models of two magnetic atom types are examined. 
By detailed investigation on hysteresis loops, multiple hysteresis behaviors are obtained for 
a given set of Hamiltonian parameters. Besides, the quantities of hysteresis characteristics as the 
hysteresis loop area, remanent magnetization, and coercive field have been investigated as functions of concentration.

\textbf{Keywords:} Hysteresis characteristics, binary alloy, effective field theory

\section{Introduction}\label{introduction}

The development and characterization of the transition metal alloys and rare earth alloys
are high interest due to the emergence of successful applications in this rapidly evolving
field \cite{ref1}. Multicritical points of rare earth alloys such as $Tb-Er$, $Dy-Er$, $Tb-Tm$, 
$Dy-Tm$ and $Gd-Er$ have been investigated in the presence of crystal field effects \cite{ref2}.
Phase diagrams of magnetic and non-magnetic transition metal alloys have been studied both
experimentally and theoretically based on Ising-type phenomenological models \cite{ref3}.

Disordered binary alloys represented by $A_c B_{1-c}$ can be modeled theoretically using well-known
Ising-like models. Binary alloy systems consisting of different spin values, such as half integer - integer spin
valued models are investigated by means of the effective field theory (EFT) \cite{ref4,ref5,ref6}, mean field theory 
(MFT) \cite{ref7,ref8,ref9} and Monte Carlo (MC) simulations \cite{ref10}. Models with half integer - half integer spins are also examined by means of MFT \cite{ref11} and also within the two frameworks EFT and
MFT \cite{ref12,ref13}. Besides, several spin systems are modeled such as $S_A=1/2$ and another component is
generalized spin-$S$ by use of MFT \cite{ref14} and within  EFT and MFT \cite{ref12}. 
Generalized of both spin variables of binary ferromagnetic alloy has been investigated with competing anisotropies 
by means of MFT \cite{ref15}. Besides, $Fe_pAl_q$ alloys \cite{ref16} and $NiBi$ alloys \cite{ref17} 
have been constructed on the Ising model within the framework of EFT. Site-diluted Ising spin model for
$Fe_{1-q}Al_q$ alloys have been examined by use of EFT \cite{ref18,ref19} and pair approximation \cite{ref20}.
The bond disordered Blume-Capel model for ternary $(Fe_{0.65}Ni_{0.35})_{1-x}Mn_x$ and $Fe_pAl_qMn_x$ 
alloys have been studied with mean-field renormalization group (MFRG) method \cite{ref21}. The Potts-like
model has been utilized to describe $Gd_{1-x}C_x$ alloy on basis of MC method \cite{ref22}.

It is crucial to emphasize that magnetic materials used in important technological applications are represented by
higher spin systems. $AB_pC_{1-p}$ ternary alloy system corresponding to the magnetic Prussian blue
analogs of \\
$(Ni_p^{II}Mn_{1-p}^{II})_{1.5}[Cr^{III}(CN)_6] \cdot zH_2O$ type consisting of 
$S_A=3/2$, $S_B=1$, $S_C=5/2$ have been investigated by use of MFT \cite{ref23} and MC \cite{ref24}.
The $(Fe_p^{II}Mn_{1-p}^{II})_{1.5}[Cr^{III}(CN)_6] \cdot nH_2O$ type analog consisting of $S_A=3/2$, $S_B=2$, $S_C=5/2$ 
has been investigated by use of MFT \cite{ref25}. Besides, different spin variables have been studied such as
$S_A=1/2$, $S_B=1$, $S_C=3/2$ \cite{ref26} and $S_A=1$, $S_B=3/2$, $S_C=1/2$ \cite{ref27} within the framework of EFT.

On the other hand, hysteresis behavior of magnetically ordered organic and molecule-based materials has been inspected extensively 
\cite{ref28,ref29}. A molecular-based magnetic material
$AFe^{II}Fe^{III}(C_2O_4)_3$ which corresponds to ferrimagnetic mixed spin-$2$ and spin-$5/2$ Ising model
on a honeycomb lattice exhibits $2S+1$ magnetic plateaus in the presence of single-ion anisotropies and at low 
temperatures. Triple and double hysteresis behaviors have been found by means of MC \cite{ref30} and EFT \cite{ref31}  for the given system.   

In Ref. \cite{ref32}, Ak{\i}nc{\i} concluded that crystal field diluted $S-1$ Blume-Capel model has double and 
triple hysteresis behavior in the ground state, at negative large values of the crystal field within the framework of
EFT. Then, generalization of these results to a spin-$S$ $(S>1)$ Blume Capel \cite{ref33} and Heisenberg models \cite{ref34} have been realized.
These models display $2S$-windowed hysteresis loop in this region. These works also focused on the difference 
between integer and half integer spin models. While half integer spin model displays central loop (which is symmetric 
about the origin), integer spin model does not exhibit central loop. 

Similar results for other systems have been reported in the literature. For instance, the binary alloy system consisting of spin-$1/2$ and
spin-$1$ could exhibit double hysteresis behavior in a concentration $0<c<0.557$ range within the framework of
EFT \cite{ref5}. The effects of the symmetric double Gaussian random field distribution on this system has also been 
investigated by use of EFT, and double hysteresis character has been obtained depending on Hamiltonian parameters \cite{ref35}.
Hysteresis behavior of quenched disordered binary alloy cylindrical nanowire consisting of the same spin set as 
the previous model has been examined by MC \cite{ref36}. The effects of the concentration and temperature on disordered 
$Fe_xAl_{1-x}$ alloys have been studied by using first-principle theory and MC simulation \cite{ref37}.
There are also studies in the literature involving the hysteresis features of higher spin models. For example, mixed
spin-$1/2$ and spin-$3/2$ Ising ferromagnetic and ferrimagnetic bilayer system on a honeycomb lattice has been studied
within EFT \cite{ref38}. It is noteworthy that, multiple hysteresis behaviors have been obtained on the mixed 
spin-$7/2$ and spin-$3/2$ Ising ferrimagnetic model by use of MC simulation \cite{ref39}. Furthermore, the dynamical hysteresis
properties of mixed spin-$3/2$ and spin-$5/2$ ferrimagnetic Ising model have been obtained by means of EFT based on
Glauber-type stochastic dynamics \cite{ref40}. Single and triple dynamical hysteresis loops have been observed 
in a spin-$1/2$ Ising bilayer system using the same method \cite{ref41}.

There are comprehensive experimental studies about magnetic alloy systems presented.  Some of them are, amorphous 
$Fe_{100-x}B_x$ alloys \cite{ref42}, $Fe$-doped $Au_xPd_{1-x}$ alloys \cite{ref43}, $Mn_xZn_{1-x}F_2$ alloys in the 
presence of random field \cite{ref44}, $Tb_xY_{1-x}$ and $Tb_xGd_{1-x}$ alloys in the presence of single-ion
anisotropy \cite{ref45}, $Fe_{100-x}Ni_x$ and $Mn_{100-x}Ni_x$ alloys \cite{ref46}, $Fe_{100-q}Al_q$ \cite{ref47} 
and $Fe_{x}Al_{1-x}Mn_x$ alloys \cite{ref48}.

In our recent work, we have investigated the hysteresis behavior of the binary 
magnetic alloy that consist of spin-1 and spin-$1/2$ atoms \cite{ref5}. The main aim of this work is to generalize the results obtained in our recent work. We want to determine the
hysteresis characteristics of the magnetic binary alloys by considering both spins which are generalized by spin-$S$ of type-$A$ 
and type-$B$ atoms consisting of different spin values. For this aim, the outline of this paper as follows:
In Sec. \ref{model} we briefly present the model and  formulation. The results and discussions are
presented in Sec. \ref{results}, and finally Sec. \ref{conclusion} contains our conclusions.

\section{Model and Formulation}\label{model}

The system consists of randomly distributed $A$ type atoms that have spin-$S_A$ and $B$ type atoms that have spin-$S_B$  within the Ising model. The concentrations of the  $A$ type atoms are denoted as $c$, and the
$B$ type atoms are denoted as $1-c$. Therefore the chemical formula is given by  $A_cB_{1-c}$. Note that, the lattice has no vacancies. 
The Hamiltonian of the binary Ising model is given by
\eq{denk1}{\mathcal{H}=-J\summ{<i,j>}{}{\paran{\xi_i \xi_j  \sigma_i \sigma_j+
\xi_i \delta_j \sigma_i s_j+
\delta_i \xi_j s_i \sigma_j+
 \delta_i \delta_j s_i s_j}}-D\summ{i}{}{\paran{\xi_i \sigma_i^2+\delta_i 
s_i^2}}-H\summ{i}{}{\paran{\xi_i \sigma_i + \delta_i s_i}},}
where $\sigma_i,s_i$ are the $z$ components of the spin-$S_A$ and spin-$S_B$ 
operators 
and they take the values
$\sigma_i=-S_A,-S_A+1,\ldots, S_A-1,S_A$ and $s_i=-S_B,-S_B+1,\ldots, S_B-1,S_B$, respectively.  
$J>0$ is the   ferromagnetic
exchange interactions between the nearest neighbor spin pairs, $D$ is
the crystal field (single ion anisotropy) and $H$ is the longitudinal magnetic field.
Here, $\xi_i=1$ means that lattice site $i$ is occupied by type-A type atoms and   $\delta_i=1$ means that lattice site $i$ is occupied by type-B type atoms. The site occupation number 
holds the relation $\xi_i+\delta_i=1$. 
The first summation in Eq.
\re{denk1} is over the nearest-neighbor pairs of spins and the other
summations are over all the lattice sites.

In a typical EFT approximation, we consider specific site (denoted by $0$) and nearest neighbors of it. All interactions of this site can be represented by $ 
\mathcal{H}_0^{A} / \mathcal{H}_0^{B}$ if the site $0$ occupied by type-A/B atoms, respectively. These terms can be treated as local fields  acting on a site $0$,

\eq{denk2}{
\mathcal{H}_0^{A}=-\xi_0\sigma_0\left[J\summ{j=1}{z}{\paran{\xi_j \sigma_j + 
\delta_j  s_j }}+H\right]-\xi_0\paran{\sigma_0}^2 D,} 

\eq{denk3}{
\mathcal{H}_0^{B}=-\delta_0 s_0\left[J\summ{\delta=1}{z}{\paran{\xi_j  \sigma_j 
+ \delta_j s_j }}+H\right]-\delta_0\paran{s_0}^2 D .} 

By defining spin-spin interaction part of these local fields as,
\eq{denk4}{E_0^{A}=
J\summ{j=1}{z}{\paran{\xi_j \sigma_j + 
\delta_j  s_j }},\quad E_0^{B}=
J\summ{\delta=1}{z}{\paran{\xi_j  \sigma_j 
+ \delta_j s_j }}}

we can write Eqs. \re{denk2} and \re{denk3} more compact form as, 
\eq{denk5}{
\mathcal{H}_0^{A}=-\xi_0\sigma_0\left[E_0^{A}+H\right]-\xi_0\paran{\sigma_0}^2 D,} 

\eq{denk6}{
\mathcal{H}_0^{B}=
-\delta_0 s_0\left[E_0^{B}+H\right]-\delta_0\paran{s_0}^2 D.} 

For obtaining magnetizations ($m_A,m_B$) and quadrupolar moments ($q_A,q_B$) for the system, we can  use the exact identities \cite{ref15a} which are given by

$$
m_A=\frac{\sanddr{\xi_0\sigma_0}}{\sandd{\xi_0}_r}=\frac{1}{\sandd{\xi_0}_r}
\sanddr{\frac{Tr_0\xi_0 \sigma_0 \exp{\paran{-\beta
\mathcal{H}_0^{A}}}}{Tr_0\exp{\paran{-\beta \mathcal{H}_0^{A}}}}},
$$

$$
q_A=\frac{\sanddr{\xi_0\sigma_0^2}}{\sandd{\xi_0}_r}=\frac{1}{\sandd{\xi_0}_r}
\sanddr{\frac{Tr_0\xi_0 \sigma_0^2 \exp{\paran{-\beta
\mathcal{H}_0^{A}}}}{Tr_0\exp{\paran{-\beta \mathcal{H}_0^{A}}}}},
$$

\eq{denk7}{m_B=\frac{\sanddr{\delta_0s_0}}{\sandd{\delta_0}_r}=\frac{1}{\sandd{
\delta_0}_r}\sanddr{\frac{Tr_0 \delta_0 s_0 \exp{\paran{-\beta
\mathcal{H}_0^{B}}}}{Tr_0\exp{\paran{-\beta \mathcal{H}_0^{B}}}}},}

$$
q_B=\frac{\sanddr{\delta_0s_0^2}}{\sandd{\delta_0}_r}=\frac{1}{\sandd{\delta_0}
_r}\sanddr{\frac{Tr_0 \delta_0 s_0^2
\exp{\paran{-\beta \mathcal{H}_0^{B}}}}{Tr_0\exp{\paran{-\beta
\mathcal{H}_0^{B}}}}}.
$$
where $Tr_0$ is the partial trace over the site
$0$, $\beta=1/\paran{k_B T}$, $k_B$ is Boltzmann constant
and $T$ is the temperature. We have two averages here, thermal averages (inner 
bracket) and random configurational averages
(bracket with subscript $r$). This last average should be taken into account to include the effect of the random distribution of the atoms in the system. 

Since all relations in Eq. \re{denk7} are in the same form, it is enough to derive one of them for obtaining the final form of the equation.  Let us choose the equation related to $m_A$ from Eq. \re{denk7}.

By writing Eq. \re{denk5} in Eq. \re{denk7} and 
performing  partial trace operations by using identity $
e^{\xi x}=\xi e^{x}+1-\xi,
$ (where $x$ is any real number and $\xi=0,1$)  we can obtain expression  in a closed form as

\eq{denk8}{
\frac{\sanddr{\xi_0 \sigma_0}}{\sandd{\xi_0}_r}=
\sanddr{f_m^A\paran{ E_0^{A}}},} where the function is given by \cite{ref16a}. All definitions of these functions will be given at the end of this section. By using differential operator technique \cite{ref17a}, Eq. \re{denk8} can be 
written as

\eq{denk9}{
\frac{\sanddr{\xi_0 \sigma_0}}{\sandd{\xi_0}_r}=\sanddrtek{e^{E_0^{A}\nabla}}f_m^A(x)|_{x=0},
} where  $\nabla$ represents the differential with respect to $x$. 
The
effect of the differential operator $\nabla$ on an arbitrary function $F$ is
given by
\eq{denk10}{\exp{\paran{a\nabla}}F\paran{x}=F\paran{x+a},} with arbitrary 
constant $a$. 

At this stage of the derivation, we have to convert the exponential operator in average braces to a polynomial form. For this aim using using approximated van der 
Waerden identities \cite{ref18a} is typical. This identity is

\eq{denk11}{
\exp{\paran{aS}}=\cosh{\paran{a\eta}}+\frac{S}{\eta} \sinh{\paran{a\eta}},
} where $\eta^2=\sandd{S^2}$ and $S$ is the spin eigenvalue. 
By using $E_0^A$ of Eq. \re{denk4} in Eq. \re{denk9} we can obtain polynomial form of the operator. Then, by using Eq. \re{denk10} with the identity $e^{\xi x}=\xi e^{x}+1-\xi,
$  we can obtain equation for $m_A$ 

\eq{denk12}{
m_A=\summ{p=0}{z}{}\summ{q=0}{z-p}{}\summ{r=0}{p}{}\summ{s=0}{z-q-r}{}\summ{t=0}{q+r}{}
C_{pqrst}(-1)^tc^{z-p}\paran{1-c}^p \paran{\frac{m_A}{\eta_A}}^q\paran{\frac{m_B}{\eta_B}}^r 
f_m^A\paran{\left[z-2s-2t\right] J,} 
}
where $\eta_A^2=q_A=\sandd{\sigma^2}$, and 
\eq{denk13}{
C_{pqrst}=\frac{1}{2^z}\komb{z}{p}\komb{z-p}{q}\komb{p}{r}\komb{z-q-r}{s}\komb{q+r}{t}.
}

By using the same steps between Eqs. \re{denk8}-\re{denk12} to other relations in Eq. \re{denk7}, we can obtain equations for other quantities as,

\eq{denk14}{
q_A=\summ{p=0}{z}{}\summ{q=0}{z-p}{}\summ{r=0}{p}{}\summ{s=0}{z-q-r}{}\summ{t=0}{q+r}{}
C_{pqrst}(-1)^tc^{z-p}\paran{1-c}^p \paran{\frac{m_A}{\eta_A}}^q\paran{\frac{m_B}{\eta_B}}^r 
f_q^A\paran{\left[z-2s-2t\right] J,} 
} 

\eq{denk15}{
m_B=\summ{p=0}{z}{}\summ{q=0}{z-p}{}\summ{r=0}{p}{}\summ{s=0}{z-q-r}{}\summ{t=0}{q+r}{}
C_{pqrst}(-1)^tc^{z-p}\paran{1-c}^p \paran{\frac{m_A}{\eta_A}}^q\paran{\frac{m_B}{\eta_B}}^r 
f_m^B\paran{\left[z-2s-2t\right] J,} 
}

\eq{denk16}{
q_B=\summ{p=0}{z}{}\summ{q=0}{z-p}{}\summ{r=0}{p}{}\summ{s=0}{z-q-r}{}\summ{t=0}{q+r}{}
C_{pqrst}(-1)^tc^{z-p}\paran{1-c}^p \paran{\frac{m_A}{\eta_A}}^q\paran{\frac{m_B}{\eta_B}}^r 
f_q^B\paran{\left[z-2s-2t\right] J.} 
} Here the functions are defined as \cite{ref16a},

\eq{denk17}{f_m^A\paran{x,H,D}=\frac{\summ{k=-S_A}{S_A}{}k\exp{\paran{\beta D 
k^2}\sinh{\left[\beta k\paran{x+H}\right]}}}{\summ{k=-S_A}{S_A}{}\exp{\paran{\beta D 
k^2}\cosh{\left[\beta k\paran{x+H}\right]}}},
}

\eq{denk18}{f_q^A\paran{x,H,D}=\frac{\summ{k=-S_A}{S_A}{}k^2\exp{\paran{\beta D 
k^2}\cosh{\left[\beta k\paran{x+H}\right]}}}{\summ{k=-S_A}{S_A}{}\exp{\paran{\beta D 
k^2}\cosh{\left[\beta k\paran{x+H}\right]}}}.
}

\eq{denk19}{f_m^B\paran{x,H,D}=\frac{\summ{k=-S_B}{S_B}{}k\exp{\paran{\beta D 
k^2}\sinh{\left[\beta k\paran{x+H}\right]}}}{\summ{k=-S_B}{S_B}{}\exp{\paran{\beta D 
k^2}\cosh{\left[\beta k\paran{x+H}\right]}}},
}

\eq{denk20}{f_q^B\paran{x,H,D}=\frac{\summ{k=-S_B}{S_B}{}k^2\exp{\paran{\beta D 
k^2}\cosh{\left[\beta k\paran{x+H}\right]}}}{\summ{k=-S_B}{S_B}{}\exp{\paran{\beta D 
k^2}\cosh{\left[\beta k\paran{x+H}\right]}}}.
}

Eqs.
\re{denk12} and \re{denk14}-\re{denk16}  constitute 
a system of coupled nonlinear equations.  The coefficients in this system are given by Eq. \re{denk13}. 
By using functions from Eqs. \re{denk17}-\re{denk19} we can solve this system by numerical procedures. 
After getting the values of $m_A,m_B,q_A,q_B$ from this solution we can calculate 
the total magnetization ($m$) and quadrupolar moment ($q$) of the system via
\eq{denk21}{
m=cm_A+\paran{1-c}m_B, \quad q=cq_A+\paran{1-c}q_B.
}

The hysteresis curves can be obtained by sweeping the magnetic field from $-H$ to $H$ and calculating magnetization in each step. The reverse sweep will yield other branch of the curve, if present.

\section{Results and Discussion}\label{results}

We consider the following scaled (dimensionless) quantities in this work

\eq{denk18}{ d=\frac{D}{J},t=\frac{k_BT}{J},h=\frac{H}{J}. }
Results have been investigated on a honeycomb lattice (i.e. $z=3$).

In this study, we will discuss the effects of the crystal field and the concentration on the hysteresis properties of 
generalized spin-S binary alloy system. Note that for different selected spin values of these atoms we can choose  both types of 
atoms as integer and half integer. We have limited this work by examining the case of $S_{A}<S_{B}$. Note also that, the 
concentration of $c$ in the case of $S_{A}<S_{B}$ corresponds to the concentration of $1-c$ in the $S_{A}>S_{B}$ case.

The hysteresis loop can be obtained by calculating the  magnetization by sweeping the magnetic field from $-h$ to $h$ direction 
and vice versa.  The system prefers the non-magnetic $s=0$ state which is the disordered phase in the ground state, 
at large negative large crystal fields, for integer spin atoms. Besides, the system is exposed to the magnetic $s=\pm 1/2$ ground 
states representing by the ordered phase at negative large crystal fields, for half integer spin atoms. When the 
magnitude of the magnetic field begins to align the spins in parallel to the field direction, the system exhibits transition from mostly occupied  ground state $s$ to
the next occupied $s+1$ ground state, at low temperatures.  If the magnetic field further increases, $s+1 \rightarrow s+2 \rightarrow ... \rightarrow S$
transitions occur, for both of integer and half integer spin values. This is the well known plateau like ground state structure. If we examine these transitions for both 
positive and negative directions of the external magnetic field, the magnetization response of the system could display
interesting hysteresis characteristics. These characteristics are the main investigation area of this work.  

\begin{figure}[h]
\epsfig{file=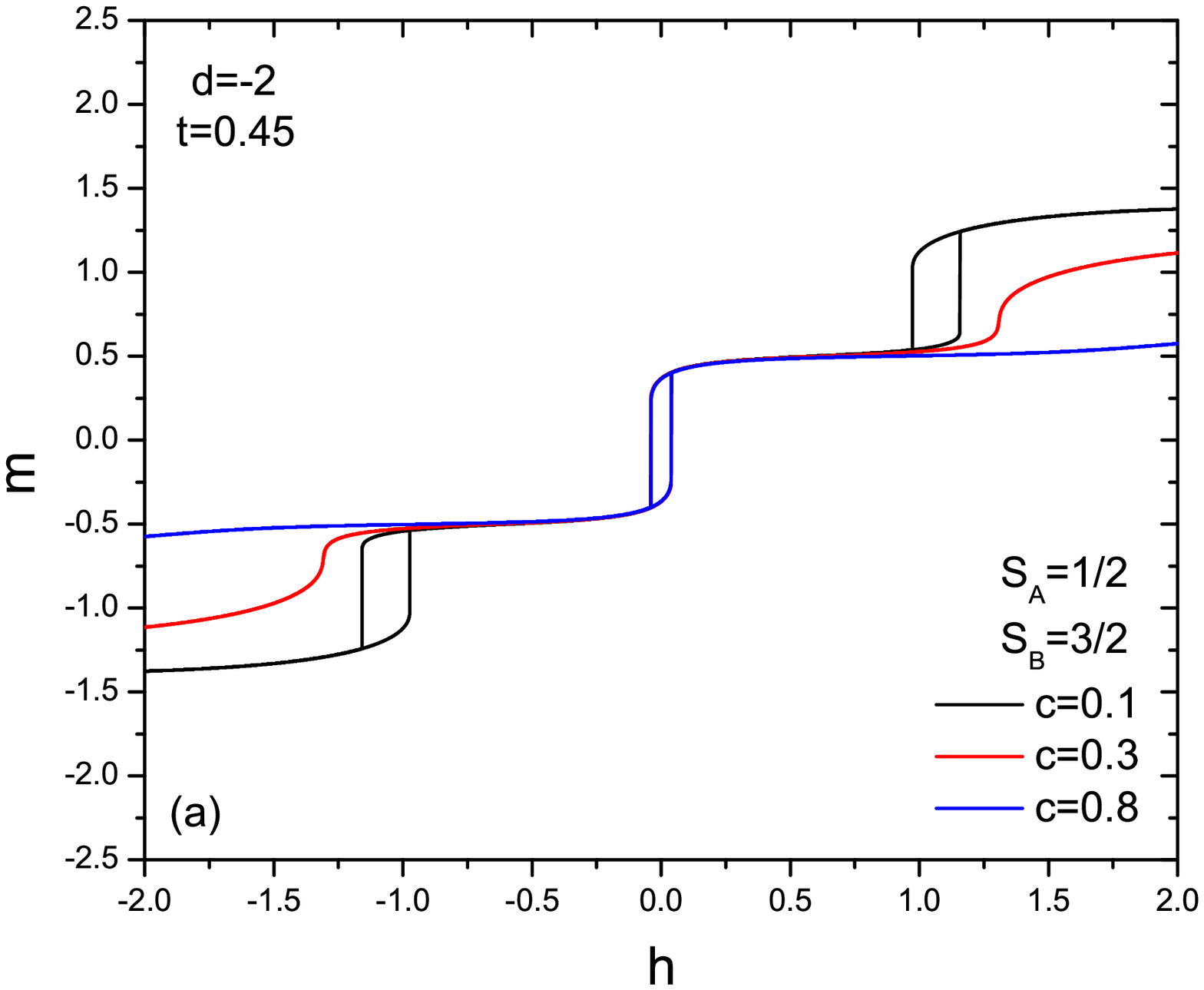, width=8cm}
\epsfig{file=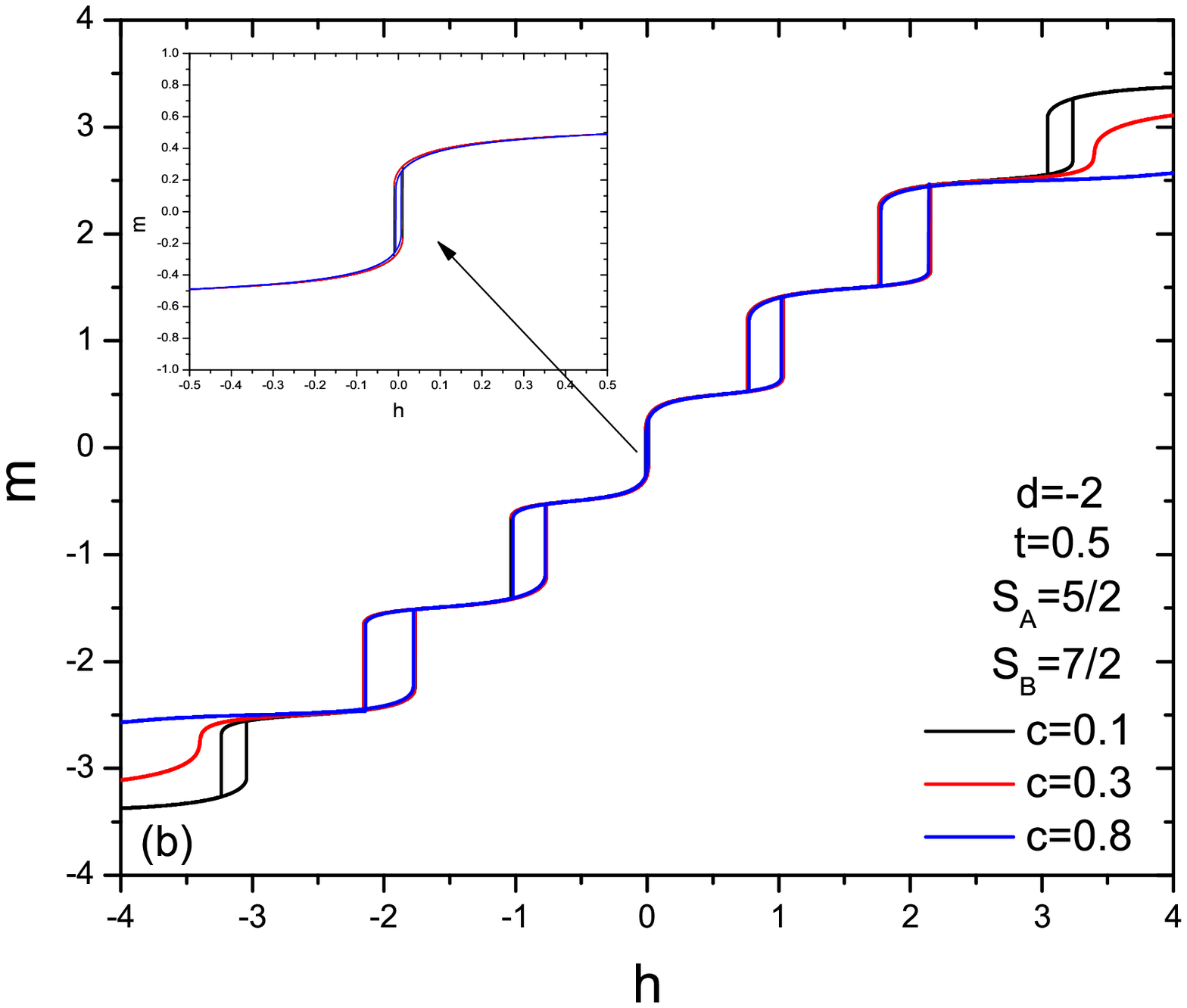, width=8cm}
\caption{Hysteresis evolutions of the binary alloy system which consist of half integer-half integer spin model such as 
(a) $S_{A}=1/2$, $S_{B}=3/2$ with temperature $t=0.45$ and (b) $S_{A}=5/2$, $S_{B}=7/2$ spin variables with $t=0.5$ for selected 
concentration values $c=0.1$, $c=0.3$ and $c=0.8$ at $d=-2$ crystal field parameter.}\label{sek1}
\end{figure}

Firstly, we choose the spin values of two different types of atoms as half integer spins. The hysteresis loops are depicted in Fig. \ref{sek1} with 
selected values of the concentrations $c=0.1$, $c=0.3$ and $c=0.8$. Spin variables of the binary alloy system are
$S_{A}=1/2$, $S_{B}=3/2$ in Fig  \ref{sek1} (a) and $S_{A}=5/2$, $S_{B}=7/2$ in Fig \ref{sek1} (b). One should notice that $c=0$ and $c=1$
cases correspond to a situation where all lattice sites are occupied by spin-$3/2$ and spin-$1/2$ atoms in Fig \ref{sek1} (a), spin-$7/2$ and spin-$5/2$ 
atoms in Fig \ref{sek1} (b), respectively. When the majority of the binary alloy system consists of $spin-3/2$ atoms, (such as $c=0.1$ in Fig. \ref{sek1} (a)),
the system displays triple hysteresis (TH) behavior for selected values of the Hamiltonian parameters $t=0.45$ and $d=-2$ (see 
Fig \ref{sek1} (a)). The central loop imply the ordered phase for this system. When we start adding more type-A atoms to the system (i.e. rising $c$),
the outer windows that appears large negative and positive values of magnetic field of hysteresis curve, disappear. The 
central loop continues with the same structure, i.e. single hysteresis (SH) is observed. When the majority of the binary alloy composed of type-A atoms  (i.e. $c=0.8$),
the ordered phase has been protected in central loop in the same way. Due to the fact that, the system prefers magnetic $s=\pm 1/2$
ground states of binary alloy which consists of half integer spins. When the system exposed to a large magnetic field, all spins
align parallel to the field. Magnetization is also greater in the system with high spin variable in large longitudinal field. For instance,
the magnetization related to $c=0.1$ is greater than the case of $c=0.8$ system for $h=2$ (see Figs. \ref{sek1} (a) and (b)). As seen in Fig. \ref{sek1} (b),
for the binary alloy system which consists of higher spin values, the number of windows of the hysteresis loops increase for a low
temperature $t=0.5$ and for crystal field parameter $d=-2$. While $c=0.1$ case exhibits seven ($2S_{B}$-windowed) windows of hysteresis (7H) loops,
$c=0.3$ and $c=0.8$ cases exhibit five ($2S_{A}$-windowed) windows of hysteresis (5H) loops . Inset of graph in Fig. \ref{sek1} (b) shows the central loop which
represent an ordered phase that exists for all concentration values. As an important result, we can emphasize that,
while the concentration decreases, two additional outermost symmetric windows of 
hysteresis loops appear. The magnetization of the system also increases for large longitudinal magnetic field, as expected. These
results are compatible with the results presented in Ref. \cite{ref33}.

\begin{figure}[h]
\epsfig{file=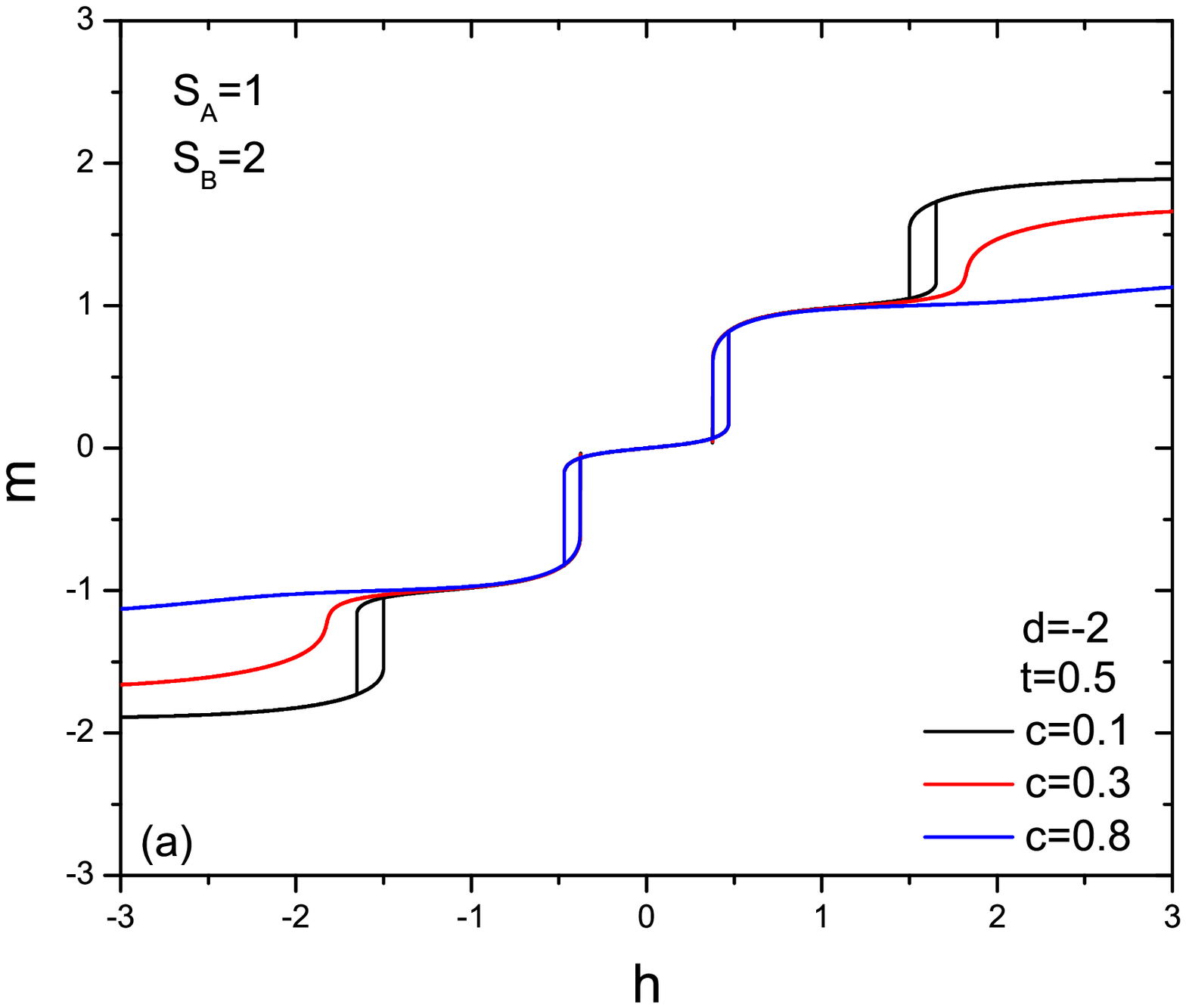, width=8cm}
\epsfig{file=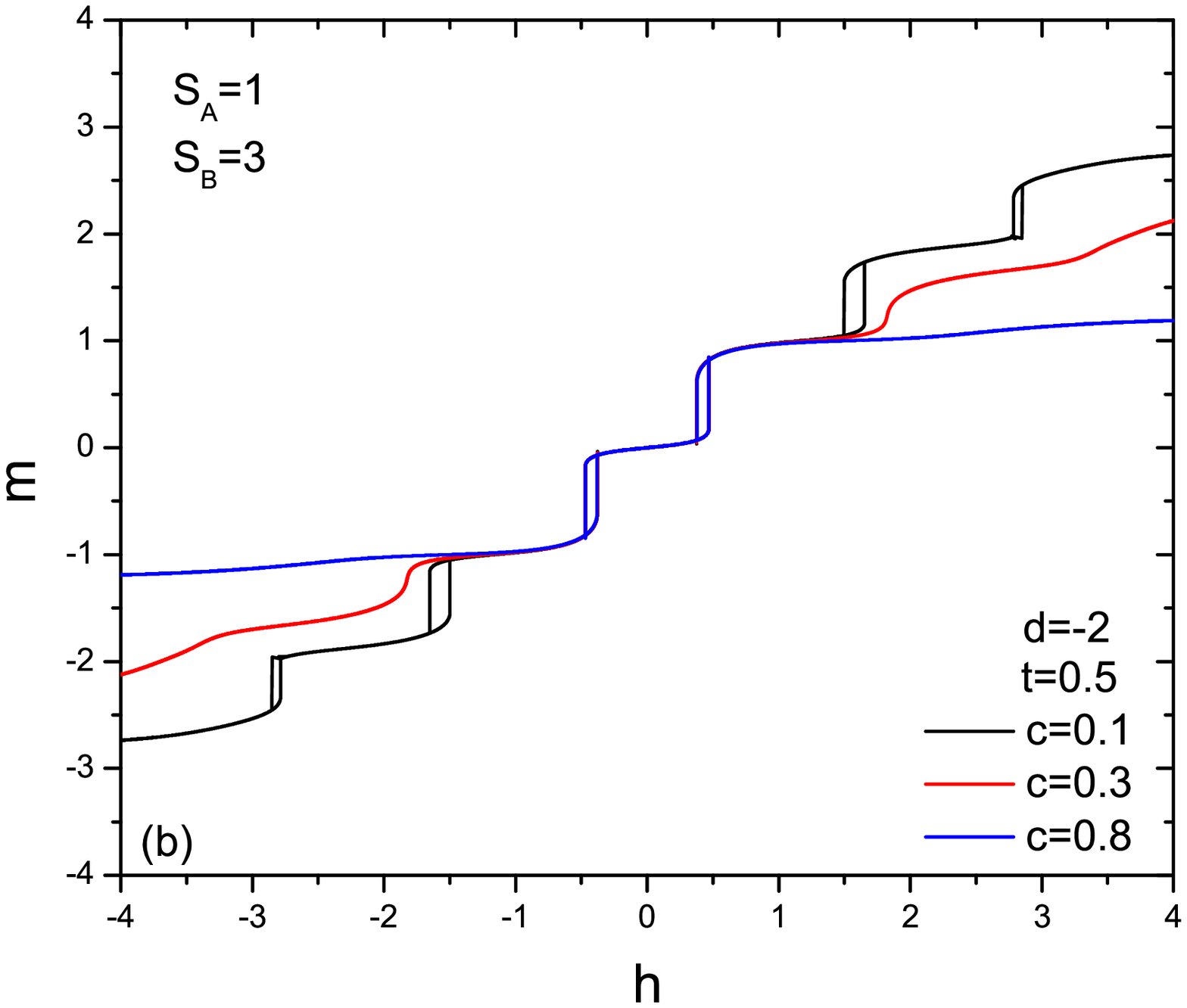, width=8cm}
\caption{Hysteresis evolutions of the binary alloy system which consist of integer-integer spin model such as (a) $S_{A}=1$, $S_{B}=2$  and
(b) $S_{A}=1$, $S_{B}=3$ spin variables for selected concentration values $c=0.1$, $c=0.3$ and $c=0.8$, with the temperature $t=0.5$ 
and $d=-2$ crystal field parameter.}\label{sek2}
\end{figure}

In Fig. \ref{sek2}, we consider a system where both of the spin variables of binary alloy are integer -values. Hysteresis properties of the system
are given for selected values of concentrations $c=0.1$, $c=0.3$ and $c=0.8$. Spin values are chosen as $S_{A}=1$ and $S_{B}=2$ in Fig.
\ref{sek2} (a) and $S_{A}=1$ and $S_{B}=3$ in Fig. \ref{sek2} (b) for $t=0.5$ and $d=-2$. In the case of $c=0.1$ the system displays four-windowed hysteresis $(4H)$
loops which have no central loop (see Fig. \ref{sek2} (a)). This means, the system has disordered phase which corresponds to the system with non-magnetic $s=0$ ground
state. 
The physical mechanism is as follows: with rising field in positive direction, $s=0$ state begins to evolve into $s=\pm 1$ states, and then these states evolves into $s=\pm 2$, with rising magnetic field. For the concentration
values of $c=0.3$ and $c=0.8$, the outer symmetric two windows of the hysteresis curve disappear and the system exhibits double hysteresis (DH) character. Magnetization decreases with increasing concentrations at large applied magnetic field. As seen in Fig \ref{sek2} (b), if we fix
the spin value of A atoms as $S_{A}=1$, and increase the spin value of B atoms such that $S_{A}=3$, then the binary alloy system with concentration $c=0.1$ exhibits 
$6$-windowed hysteresis $(6H)$ loops, without central loop. A considerable noteworthy result is that the system displays DH behavior for $c=0.3$ and $c=0.8$, due to the higher spin value of B atoms. The system eliminates the
outermost four symmetric loops to adapt from the $6$-windowed hysteresis  loop behavior to the DH loop behavior. Regardless of
the difference between the spin values of the binary alloys, the number of outer windows that disappeared will be twice that 
difference for concentration that close to the small spin value. To clarify, difference between the spin 
values of type-A and type-B atoms are $1$ (2), the number of outer windows that disappear will be two (four) with increasing components of A 
atoms in Fig \ref{sek2} (a) (Fig. \ref{sek2} (b)), respectively.

\begin{figure}[h]
\centering
\epsfig{file=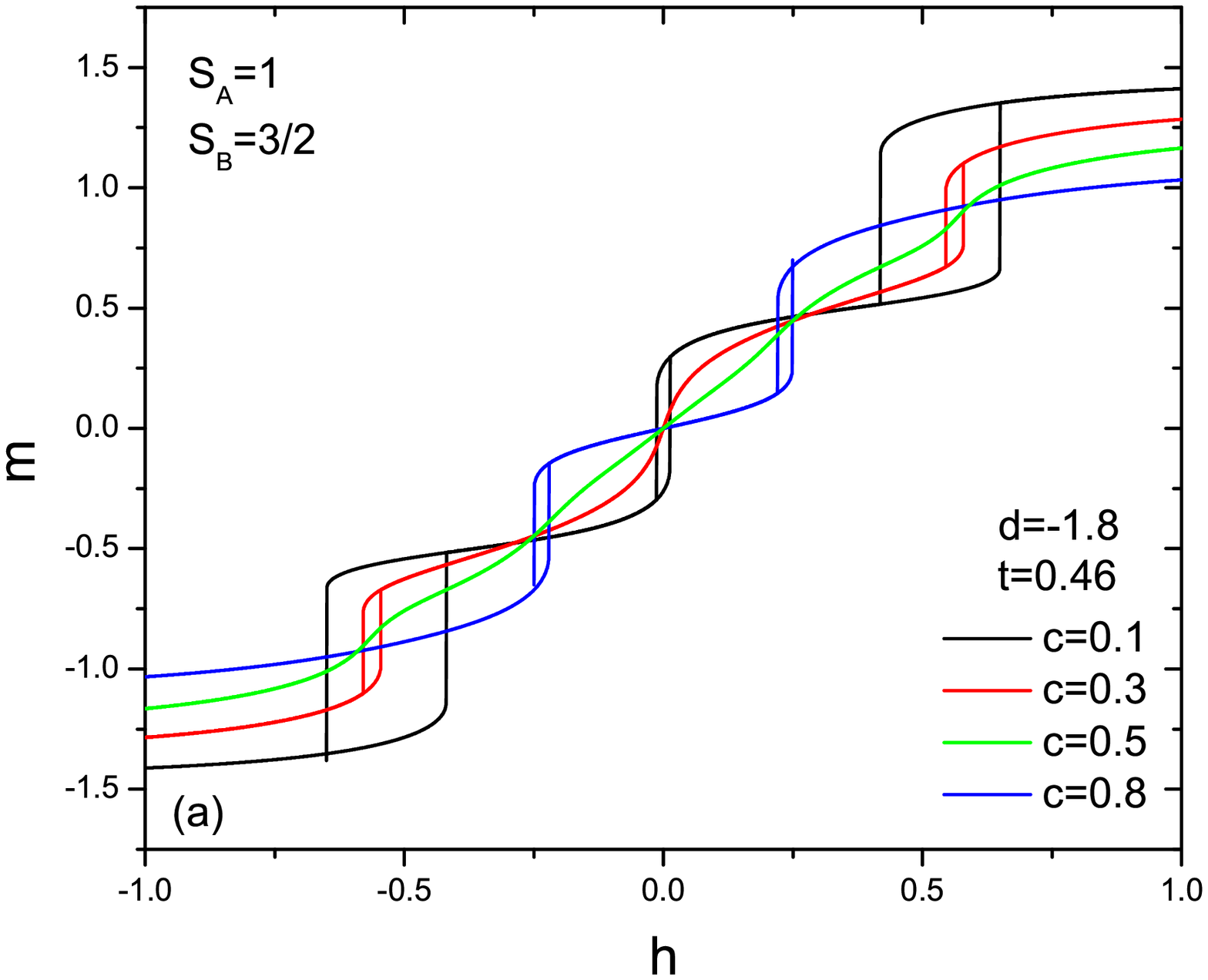, width=9cm}
\epsfig{file=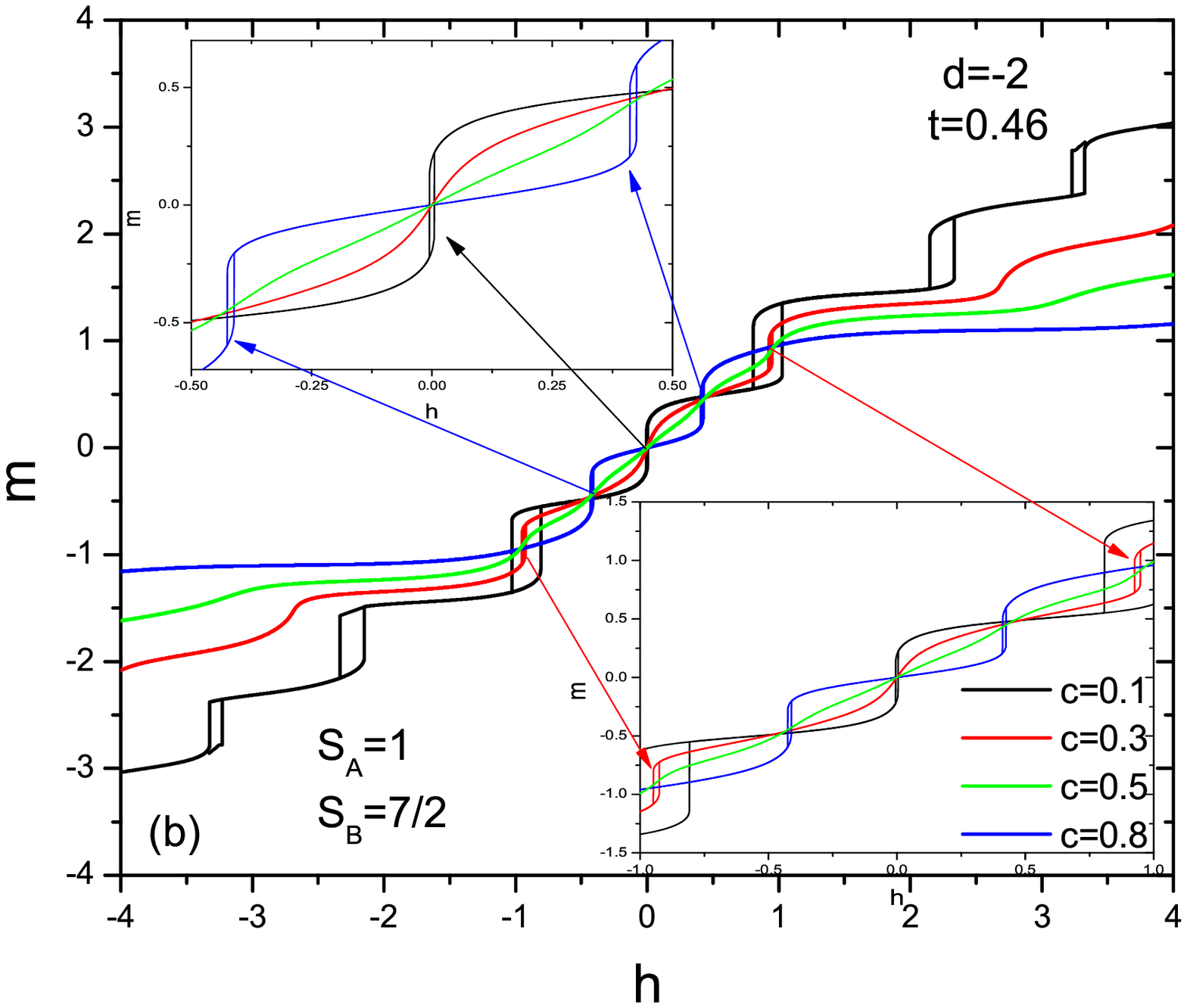, width=9cm}
\epsfig{file=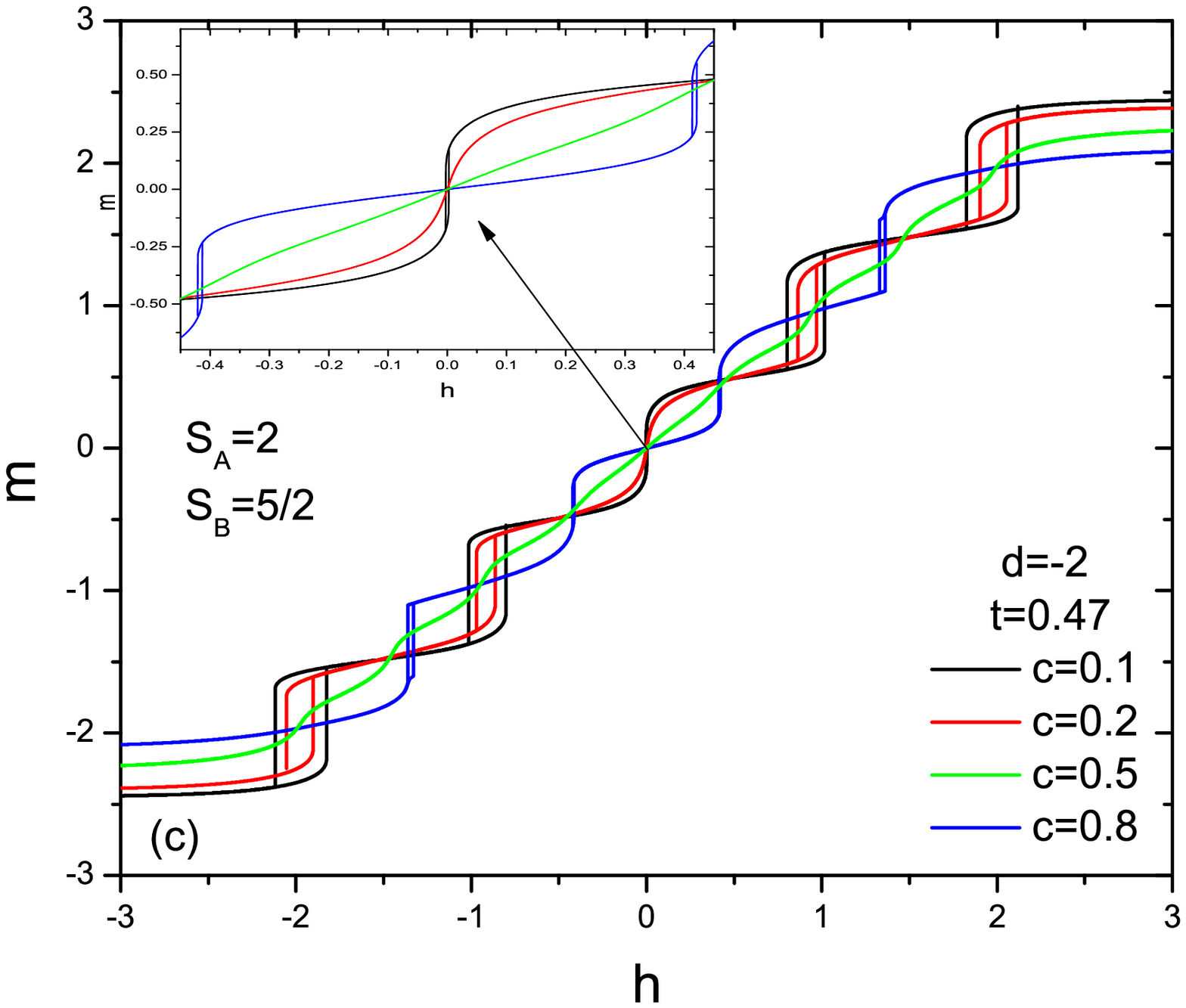, width=9cm}
\caption{Hysteresis evolutions of the binary alloy system which consists of integer- half integer spin model such as 
(a) $S_{A}=1$, $S_{B}=3/2$, (b) $S_{A}=1$, $S_{B}=7/2$ and (c) $S_{A}=2$, $S_{B}=5/2$ spin variables for a given set of 
Hamiltonian parameters.}\label{sek3}
\end{figure}

In the previous examinations, the binary alloy system had only one of the ordered (half integer spin valued binary alloys) or
disordered (integer spin valued binary alloys) phases at low temperatures and large negative crystal field values. To find out
which phase does the system exhibit in binary alloys consisting of spin values integer and half integer, we choose spin 
value of A atoms as integer and spin value of B atoms as half integer in Fig \ref{sek3}. Hysteresis behaviors of binary alloy consisting
of (a) $S_{A}=1$, $S_{B}=3/2$, (b) $S_{A}=1$, $S_{B}=7/2$ and (c) $S_{A}=2$, $S_{B}=5/2$ spin values have quite striking results. 
In figure \ref{sek3} (a), hysteresis curves are shown for the values of the concentration $c=0.1$, $c=0.3$, $c=0.5$ and $c=0.8$ for $t=0.46$ 
and $d=-1.8$. For the case of $c=0.1$, namely majority of the lattice sites consist of type B atoms, the system has TH
behavior with central loop (see the curves related to the $c=0.1$ in Figs. \ref{sek3} (a), (b) and (c)). For the concentration value $c=0.3$, the central loop disappears and the outer symmetric two
loops become narrower. Since the central loop is lost, the system should exhibit a phase transition in this concentration ranges.
For the value of $c=0.3$, the system does not have a magnetic ground state at low magnetic field values. As the applied magnetic field
increases, smaller magnetic field induces the transition between the ground states of the system. Contrary 
to the common belief, the difference between the new ground states dominated by the B atoms are less than one. The transition from TH to DH can be arranged systematically as follows: 
The number of windows that disappeared will be two times of difference 
between two spin values of type A and type B atoms, i.e. $2(S_{B}-S_{A})$ windows disappear. The difference between $S_{A}=1$ and $S_{B}=3/2$ 
spins are $1/2$, so one loop disappears, which is the central loop. For concentration value of $c=0.5$, the system exhibits paramagnetic hysteresis 
(PH) which prefers $s=0$ state (see the curves for $c=0.5$ in Fig. \ref{sek3} (a), (b) and (c)). When the concentration value increases to $c=0.8$, due to the spin value of A atoms, $2S_{A}$-windowed (namely DH) behavior appears. Depending on the concentration of the binary alloy, DH for $c=0.3$ and
$c=0.8$ are different from each other due to the difference between their ground states.
In this case,  $TH-DH-PH-DH$ hysteresis behaviors occur. The first one (TH)
corresponds to the ordered phase and the others (DH, PH and other DH) are disordered phase. Phase transition between two phases causes $2(S_{B}-S_{A})=1$
windows to disappear. In Fig \ref{sek3} (b) Hamiltonian parameters are set as $t=0.46$ and $d=-2$ for various concentrations. For $c=0.1$ system
exhibits $7$-windowed hysteresis loops with central loop which is demonstrated in the upper inset. For the value of $c=0.3$, the system exhibits
DH behavior which is demonstrated in lower inset. The observed DH loops are also narrower than
the concentric loops. So, transition from ordered to a disordered phase occurs between these concentration ranges. As
the concentration of type-A atoms increases in the system, we observe that some loops 
disappear according to the new ground states, which are dominated by the B atoms. From the general result mentioned above, $2(S_{B}-S_{A})=5$ windows should disappear.
Four of them are the outermost symmetric windows and the other is the central loop as
can be seen in Fig \ref{sek3} (b). For the value of $c=0.5$, the system exhibits disordered phase and then for $c=0.8$ the innermost
double symmetric loops appear, which are also demonstrated in upper inset of Fig. \ref{sek3} (b). For alloys consisting
mostly of  atoms type-A, the ground state will be dominated by  atoms type-A, so it has different center from 
the other DH loops. In brief, $7H-DH-PH-DH$ hysteresis 
behaviors are observed as the concentration increases. In Fig \ref{sek3} (c), the system exhibits $5H-4H-PH-4H$ hysteresis
behaviors (while the concentration increases) for $t=0.47$ and $d=-2$, respectively. Transition from the ordered phase ($(5H)$ windowed loop) to the disordered phase
($(4H)$ windowed loop) causes $2(S_{B}-S_{A})=1$ window disappears, which is the central loop. The other $4$-windowed loops get narrower. 
For the value of $c=0.5$ system has paramagnetic hysteresis (PH) loop. It is worth to emphasize here that, for $c=0.8$,  the innermost 
$4$-windowed hysteresis loops (which have different centers from the other $4$-windowed loops) appear. 

When we generalized integer-half integer results, for low
temperatures and negative large crystal fields, $2S_{B}-2S_{A}-PH-2S_{A}$-windowed hysteresis loops are observed while the concentration increases for 
$S_{A}<S_{B}$. The first one is the ordered phase and the other three are in disordered phase. 
The transition from $2S_{B}$ to $2S_{A}$-windowed loop causes
$2(S_{B}-S_{A})$ disappearing windows. The last $2S_{A}$-windowed hysteresis appears as innermost loops due to the case of $S_{A}<S_{B}$.

\begin{figure}[h]
\centering
\epsfig{file=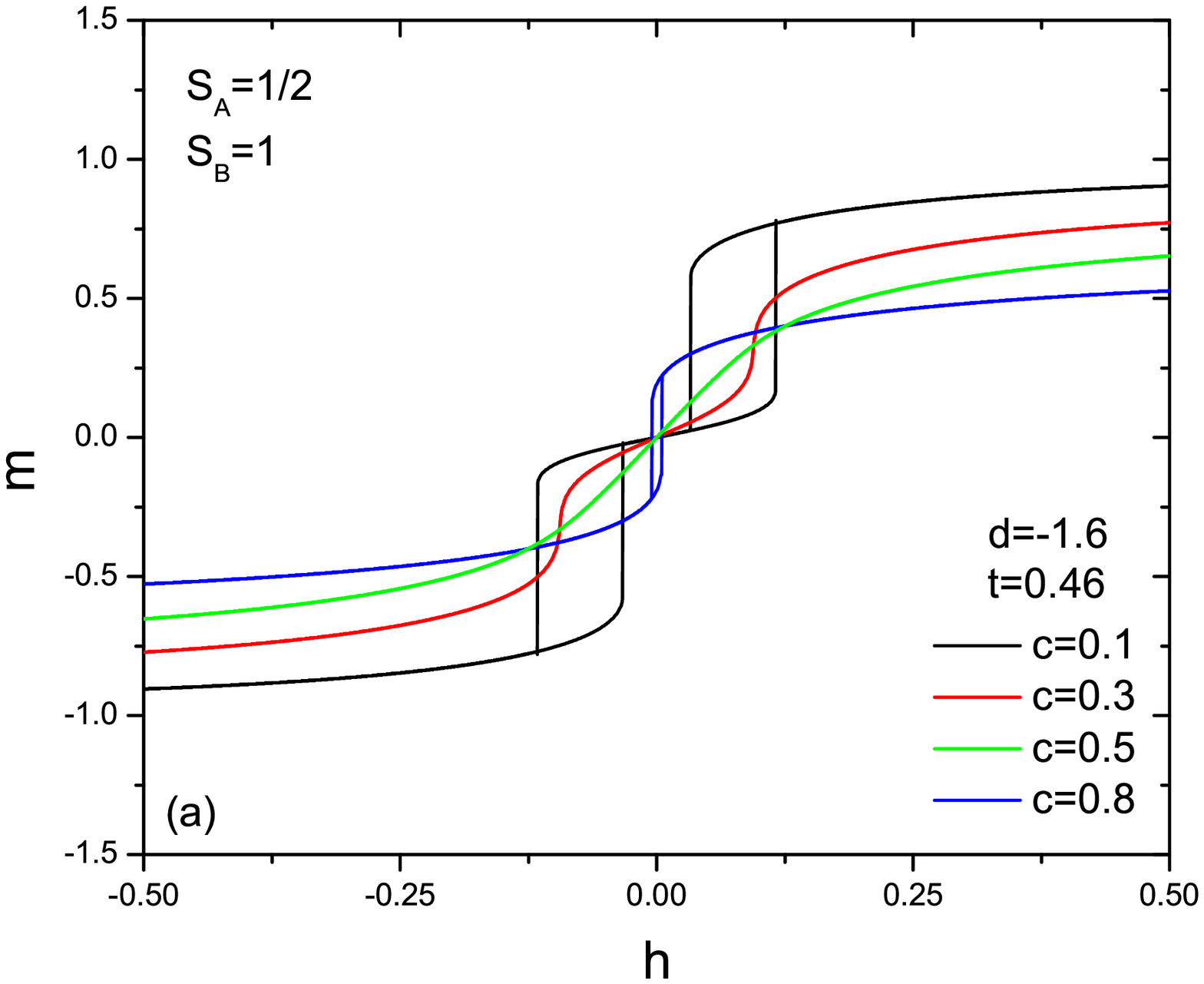, width=9cm}
\epsfig{file=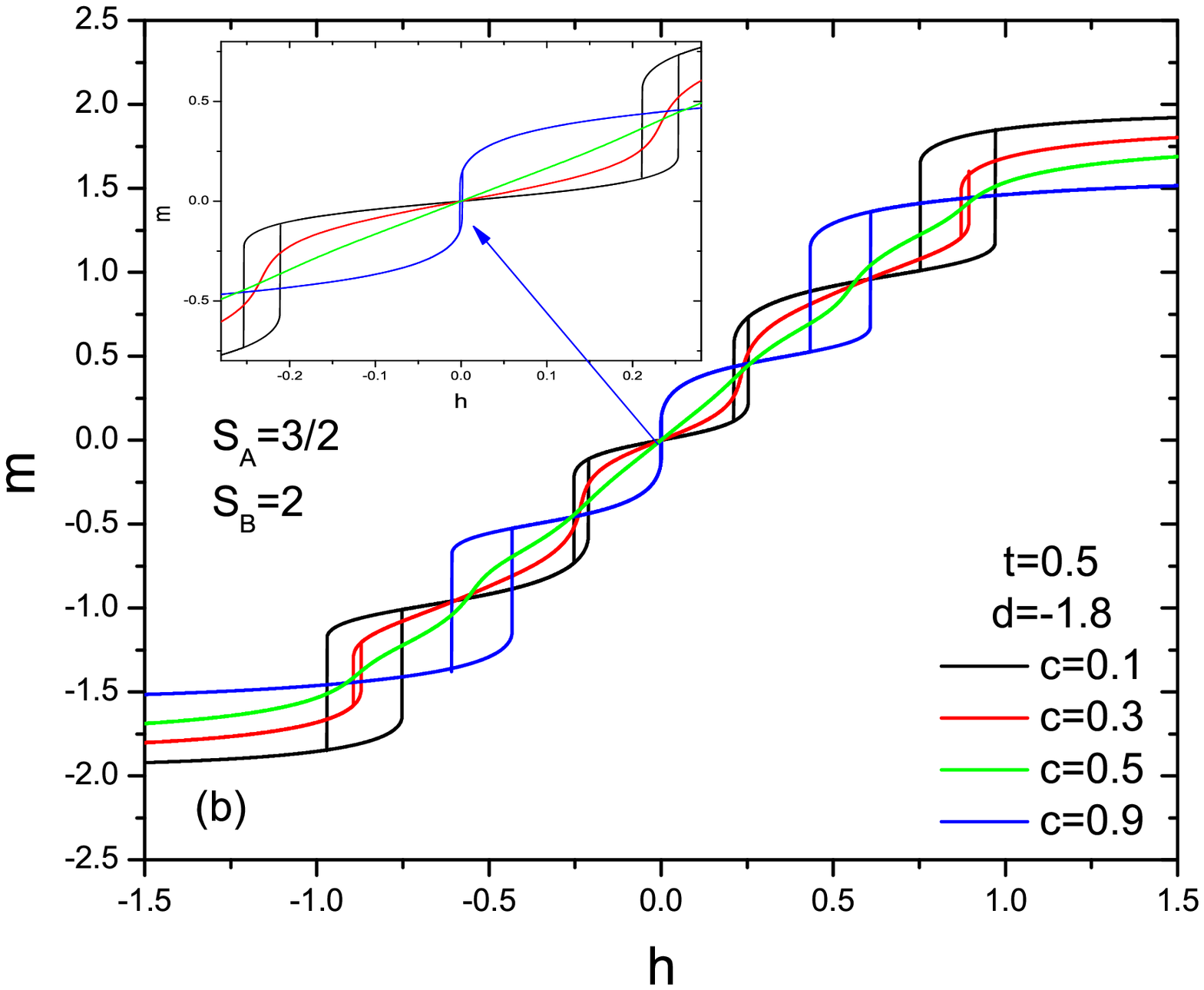, width=9cm}
\epsfig{file=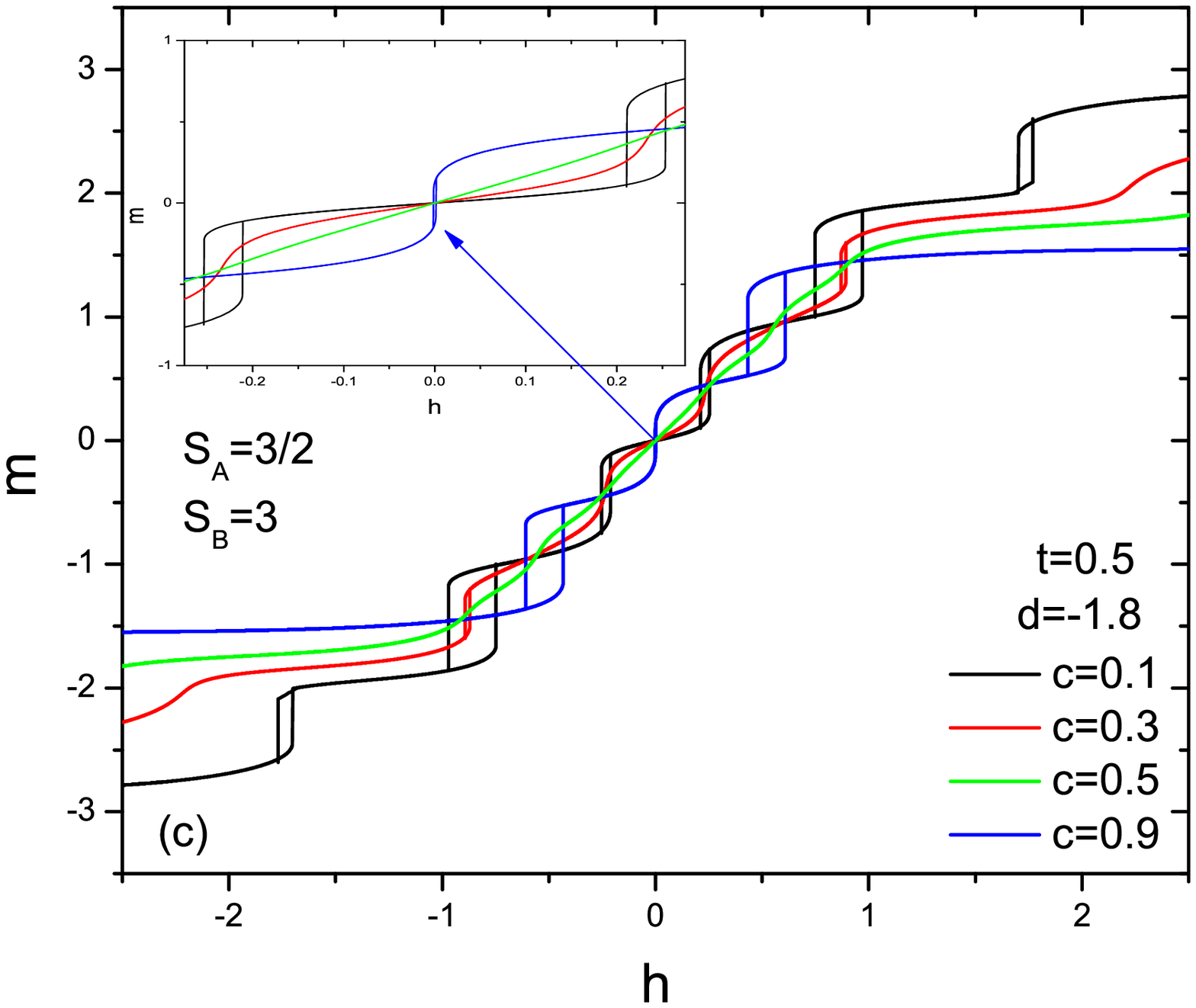, width=9cm}
\caption{Hysteresis evolutions of the binary alloy system which consists of half integer-integer spin model such as (a) $S_{A}=1/2$, $S_{B}=1$, (b) $S_{A}=3/2$, $S_{B}=2$ 
and (c) $S_{A}=3/2$, $S_{B}=3$ spin variables for a given set of Hamiltonian parameters.}\label{sek4}
\end{figure}

 In Fig \ref{sek4}, we have investigated binary alloy model consisting of half integer - integer spins such as (a) $S_{A}=1/2$, $S_{B}=1$, (b) $S_{A}=3/2$, 
$S_{B}=2$ and (c) $S_{A}=3/2$, $S_{B}=3$. In Fig \ref{sek4} (a), the system exhibits $DH-PH-PH-SH$ hysteresis behaviors for the values of  $t=0.46$ and $d=-1.6$. 
In Fig \ref{sek4} (b), $4H-2H-PH-3H$ hysteresis behaviors are obtained for $t=0.5$ and $d=-1.8$. In Fig \ref{sek4} (c), $6H-2H-PH-3H$ hysteresis behaviors are observed for $t=0.5$ and $d=-1.8$. 
Firstly, $2S_{B}$-windowed hysteresis loops are observed for lower concentration values, as expected. 
If the concentration of the system is increase from $c=0.1$ to $c=0.3$, the innermost DH disappears due
to the new ground states dominated by  atoms of type-B (see Fig. \ref{sek4} (b)). 
The system exhibits paramagnetic hysteresis behavior for $c=0.5$. If the concentration of the system is
increases from $c=0.5$ to $c=0.8$, the innermost $2S_{A}$-windowed hysteresis appears with a central loop. The first three hysteresis behaviors ($4H-2H-PH$) correspond to disordered paramagnetic
phase and the last one is ($3H$) the ordered phase. The phase transition occurs between these two phases.

\begin{figure}[h]
\epsfig{file=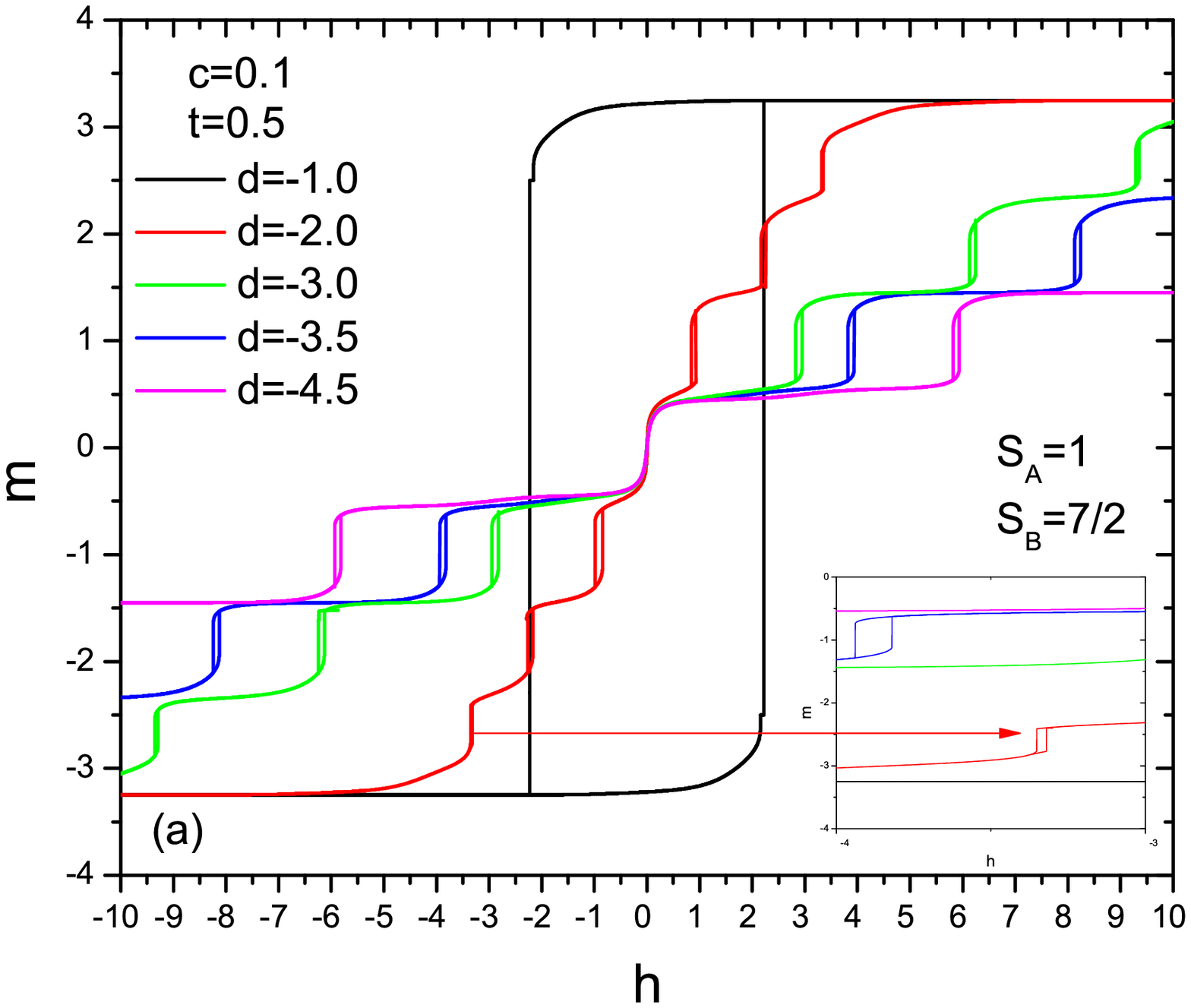, width=8cm}
\epsfig{file=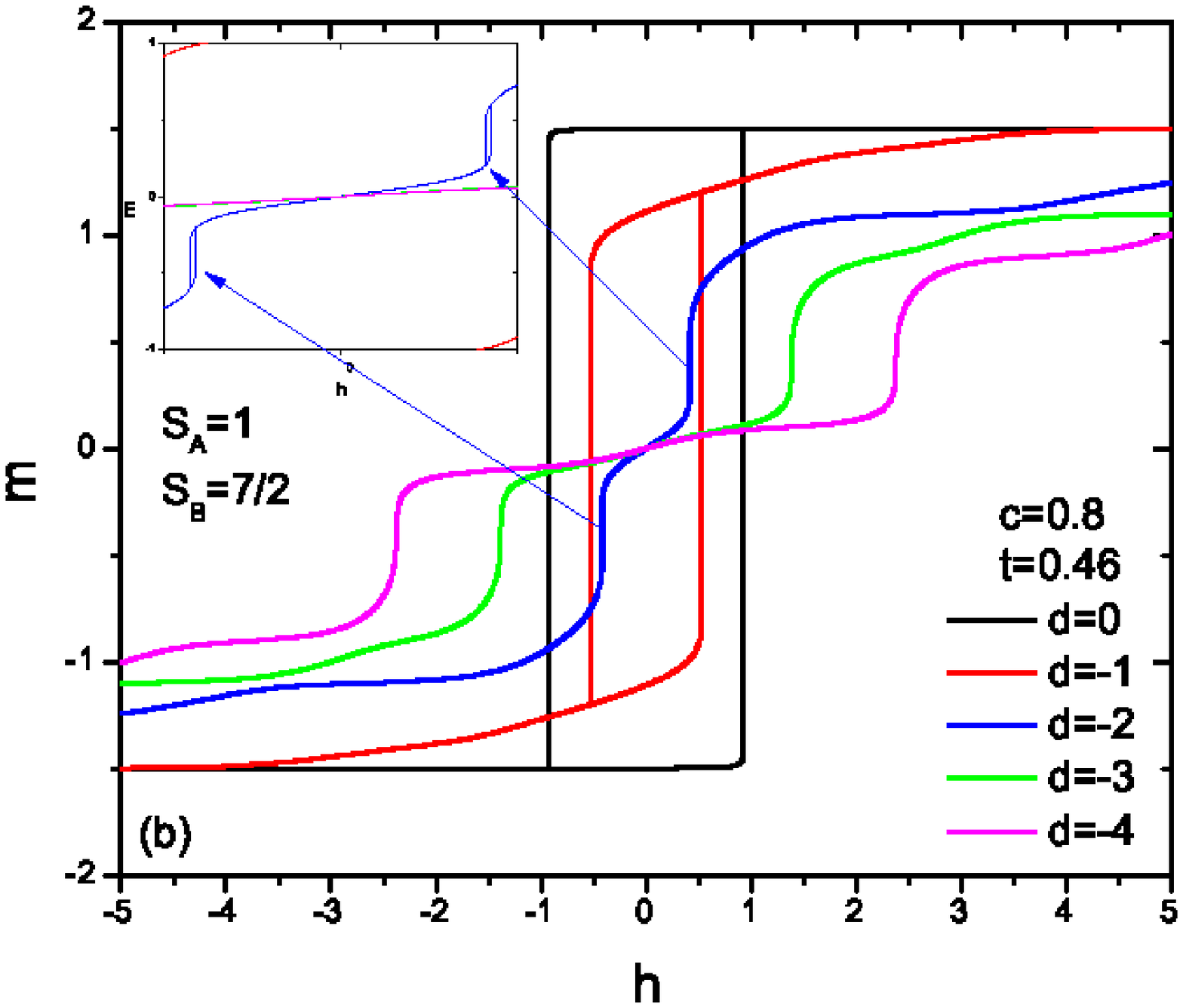, width=8cm}
\caption{Hysteresis evolutions of the binary alloy system which consists of integer- half integer spin model such as 
$S_{A}=1$, $S_{B}=7/2$ for the concentrations (a) $c=0.1$ and (b) $c=0.8$ for a given set of 
Hamiltonian parameters.}\label{sek5}
\end{figure}

\begin{figure}[h]
\epsfig{file=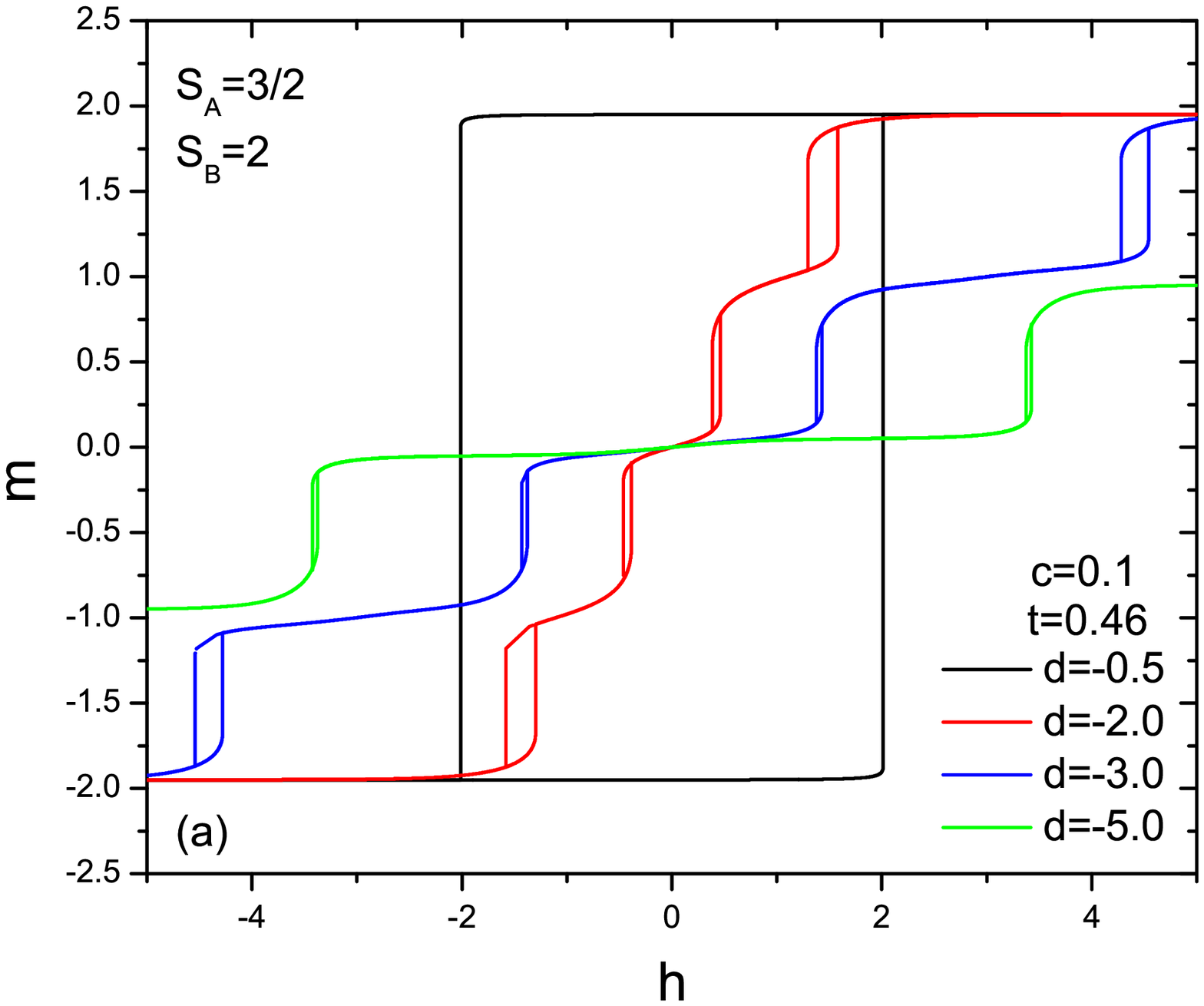, width=8cm}
\epsfig{file=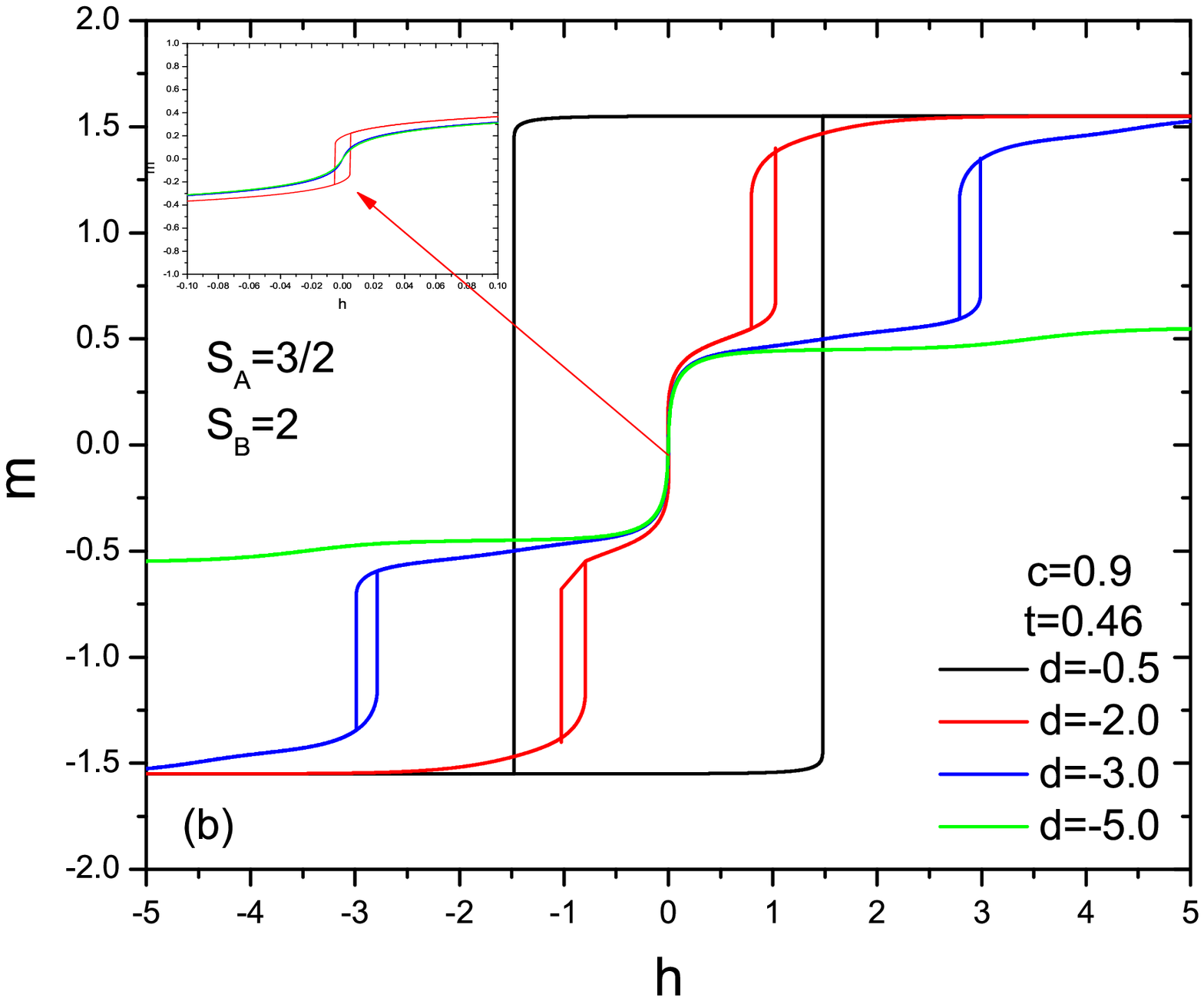, width=8cm}
\caption{Hysteresis evolutions of the binary alloy system which consists of half integer-integer spin model such as 
$S_{A}=3/2$, $S_{B}=2$ for the concentrations (a) $c=0.1$ and (b) $c=0.9$ for a given set of 
Hamiltonian parameters.}\label{sek6}
\end{figure}

The effects of crystal field parameter are examined for the binary alloys consisting of integer-half integer spin model as,
can be seen in Fig \ref{sek5}. Spin values of the system are selected as $S_{A}=1$, $S_{B}=7/2$ and for the parameters $c=0.1$ and $t=0.5$ in
Fig. \ref{sek5} (a), $c=0.8$ and $t=0.46$ in Fig. \ref{sek5} (b). As an evolution of the hysteresis loops with changing crystal field,  $SH-6H-6H-4H-2H$ is observed
 (see Fig. \ref{sek5}  (a)). We concluded by comparison of Fig. \ref{sek3}  (b) and Fig. \ref{sek5}  (a) from which we see that, by increasing the temperature, the central loop
disappears for the value of $d=-2$. The system exhibits SH behavior, which is represented by ordered phase for $d=-1$, and disordered
phase for lower values of the crystal field. As the crystal field increases to a negative large values, the outermost symmetric
loops gradually disappear  and the windows start to be separated from each other. As seen in Fig. \ref{sek5}  (b) $SH-SH-DH-PH-PH$ hysteresis behaviors are observed
for given values of the crystal field. Phase transition occurs when the  crystal field parameter changes from $d=-1$ to $d=-2$.
When the system consists of mostly type-A atoms,
multiple hysteresis behavior is replaced by SH and DH behaviors. 
In Fig. \ref{sek6}, we have investigated the effects of crystal field on binary alloy consisting of
half integer-integer spins at the temperature $t=0.46$. For the spin values of $S_{A}=3/2$, $S_{B}=2$, the system displays $SH-4H-4H-2H$ hysteresis behaviors for $c=0.1$ and  
for selected crystal field parameters (see Fig. \ref{sek6} (a)). The system exhibits a phase transition with increasing crystal field 
values, in the region between $d=-0.5$ and $d=-2$.
Hysteresis behaviors which are observed as $SH-3H-DH-PH$ for $c=0.9$, respectively can be seen in Fig. \ref{sek6} (b). The central loop disappears when
passing from $3H$ to $DH$ behavior, and phase transition occurs in this range.

In order to investigate the evolution of the hysteresis properties, we describe the quantities hysteresis loop area (HLA),
remanent magnetization (RM) and coercive field (CF). HLA is defined as the area covered by hysteresis loop in $(m,h)$ plane, and
it describes the loss of energy due to the hysteresis. The RM is residual magnetization in the system after an applied magnetic field
is removed. CF is the intensity of the magnetic field needed to change the sign of the magnetization. 

\begin{figure}[h]
\epsfig{file=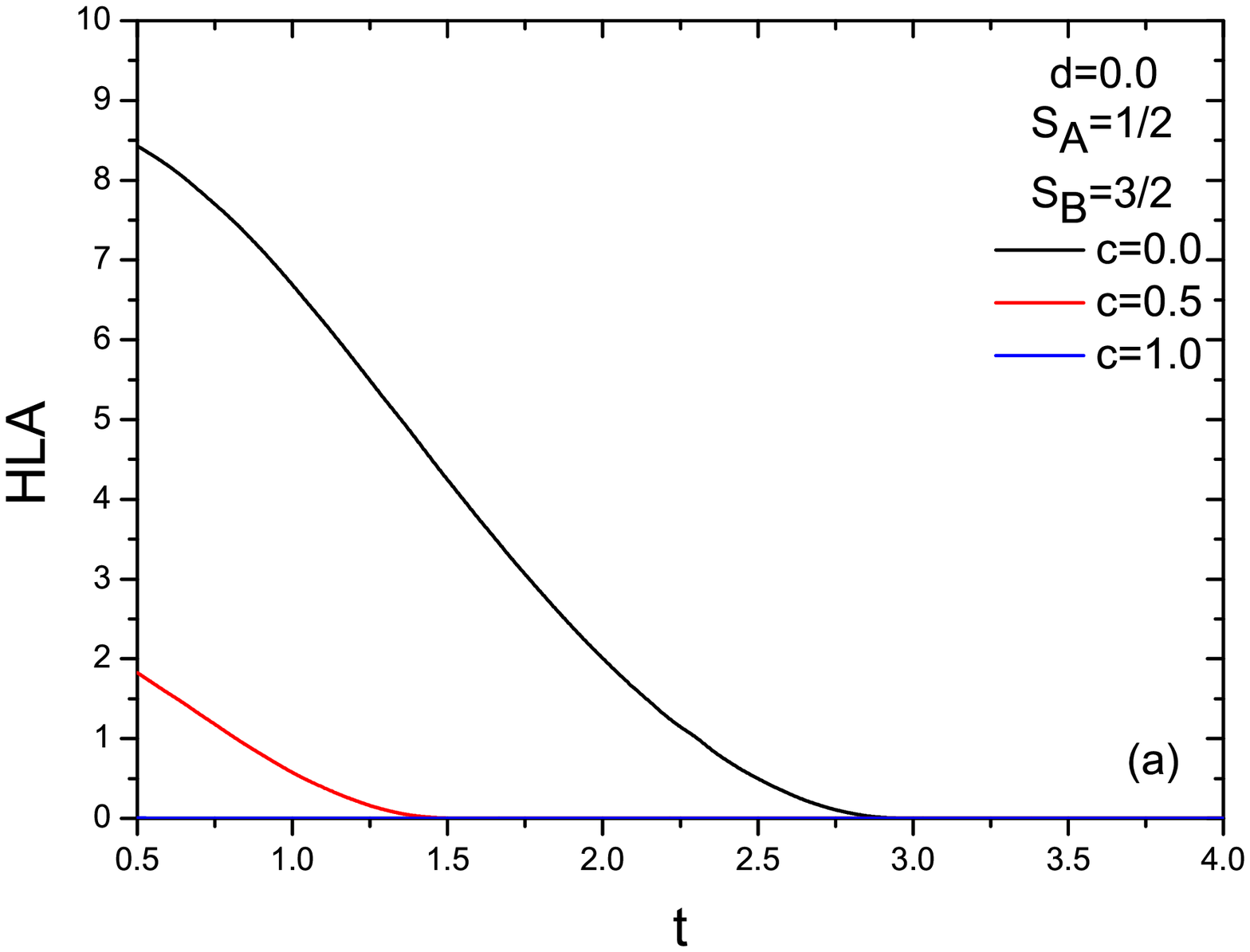, width=8cm}
\epsfig{file=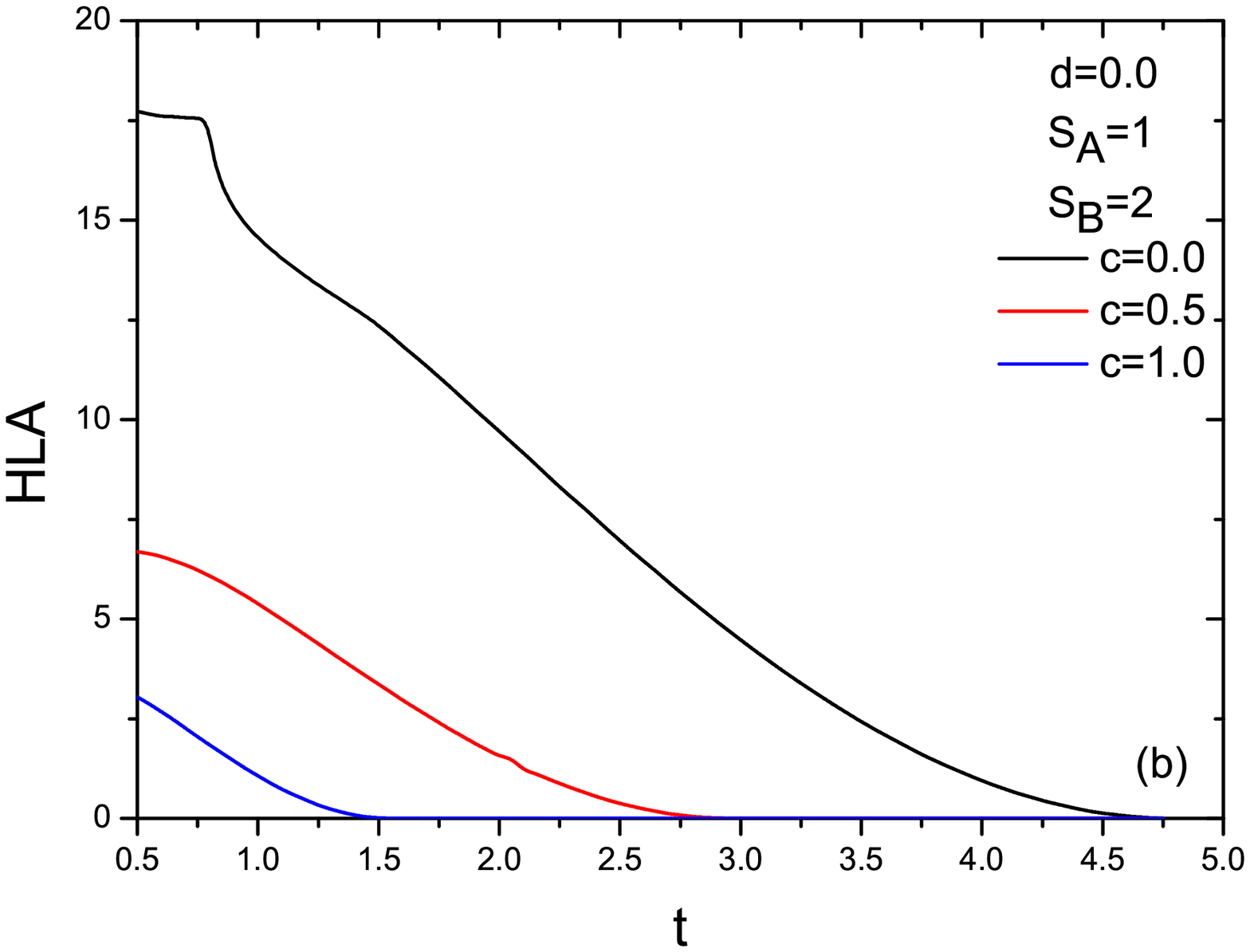, width=8cm}
\epsfig{file=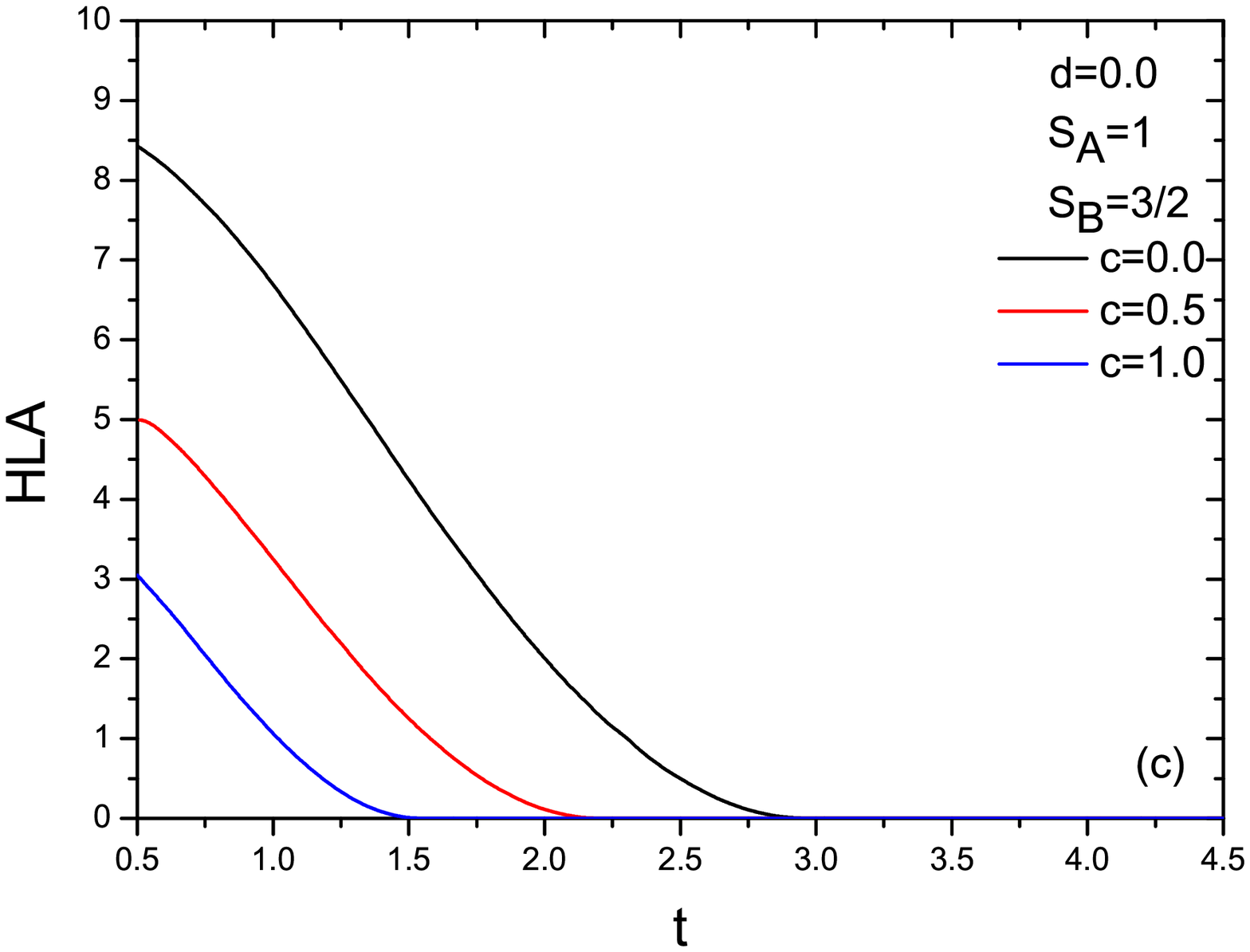, width=8cm}
\epsfig{file=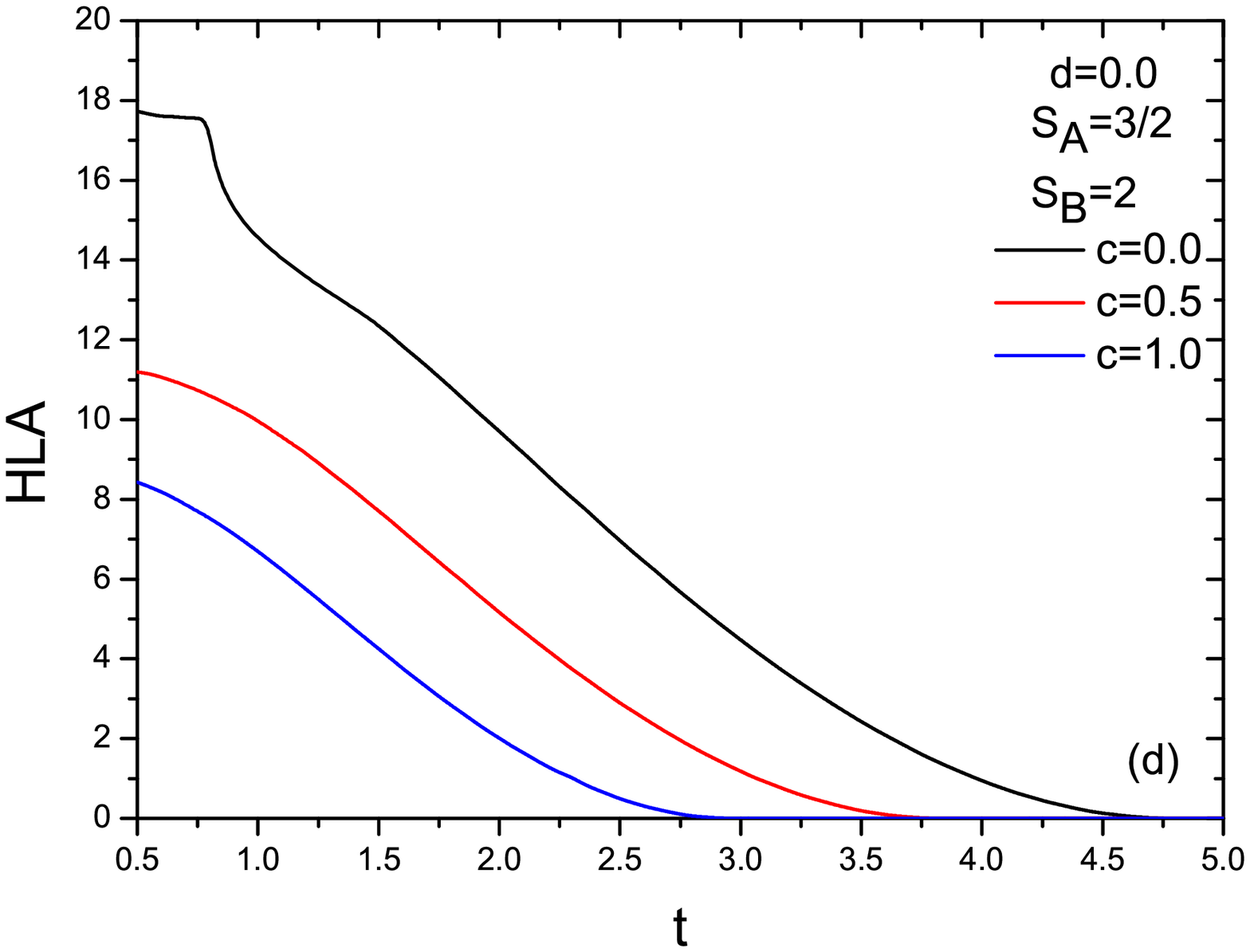, width=8cm}
\caption{Variation of HLA with the temperature of the binary alloy system which consists of 
(a) $S_{A}=1/2$, $S_{B}=3/2$, (b) $S_{A}=1$, $S_{B}=2$, (c) $S_{A}=1$, $S_{B}=3/2$
and (d) $S_{A}=3/2$, $S_{B}=2$ spin variables for selected values of concentrations $c=0.0$, $c=0.5$ and $c=1.0$ at 
$d=0$ crystal field parameter.}\label{sek7}
\end{figure}

\begin{figure}[h]
\epsfig{file=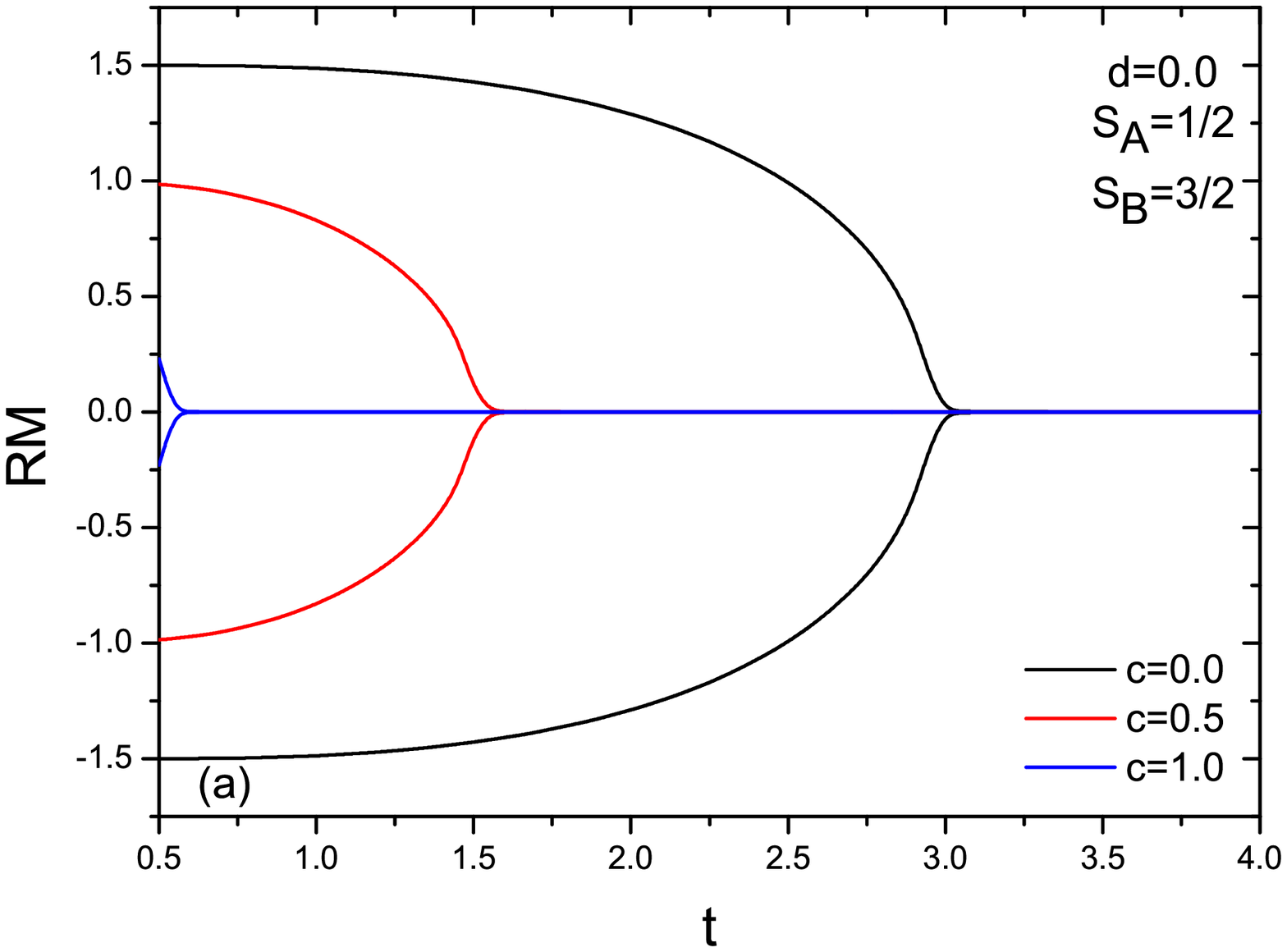, width=8cm}
\epsfig{file=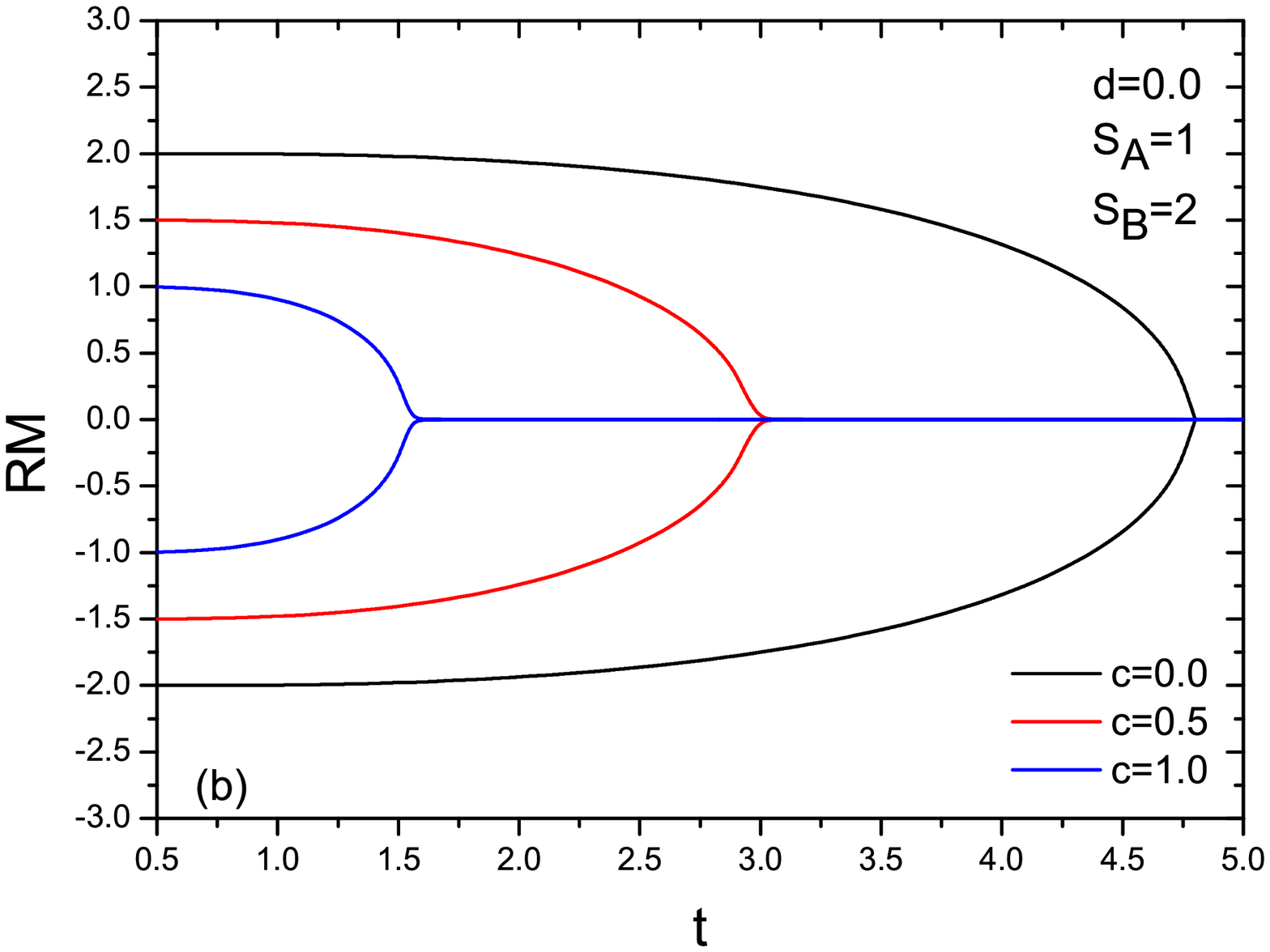, width=8cm}
\epsfig{file=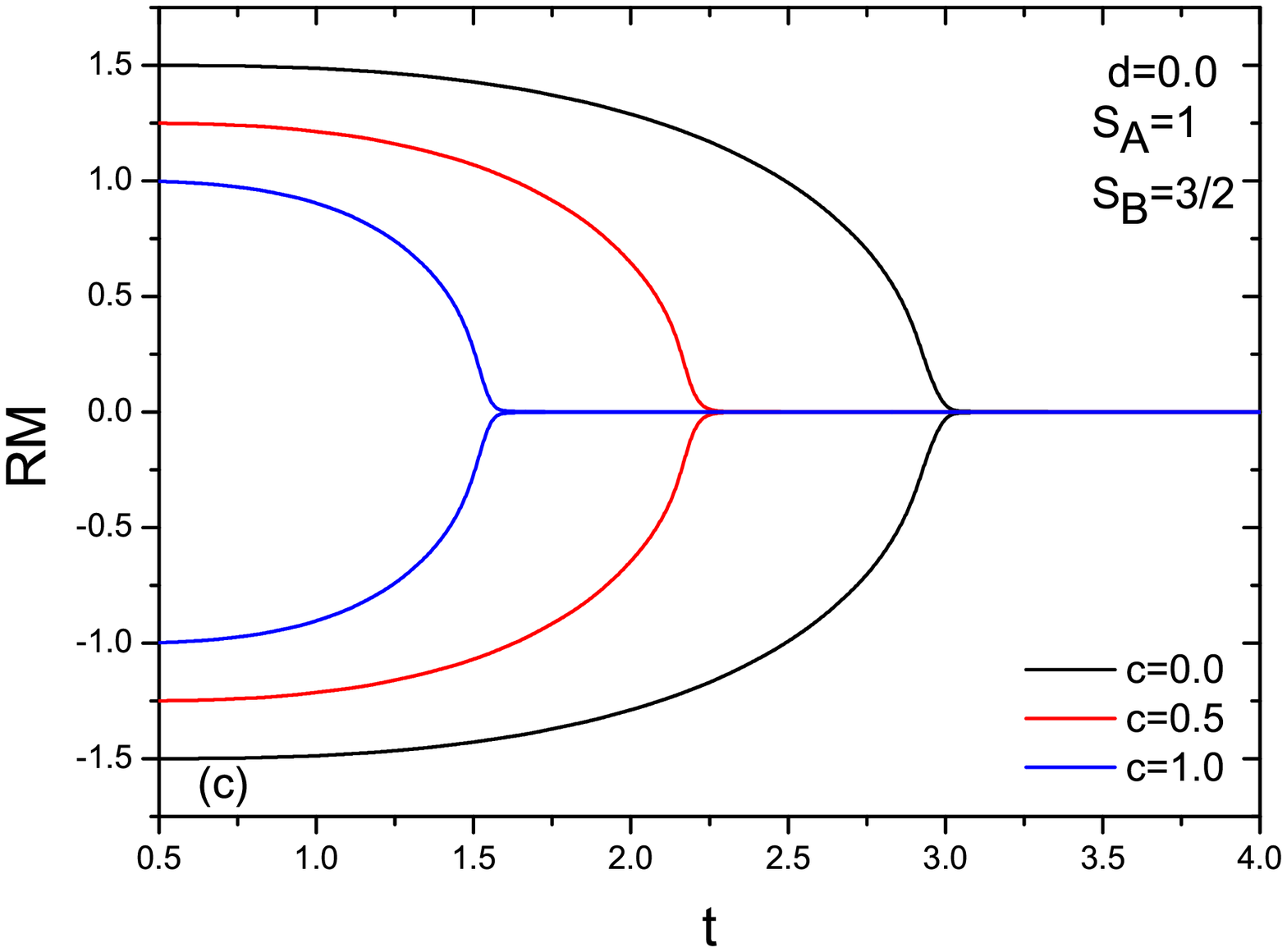, width=8cm}
\epsfig{file=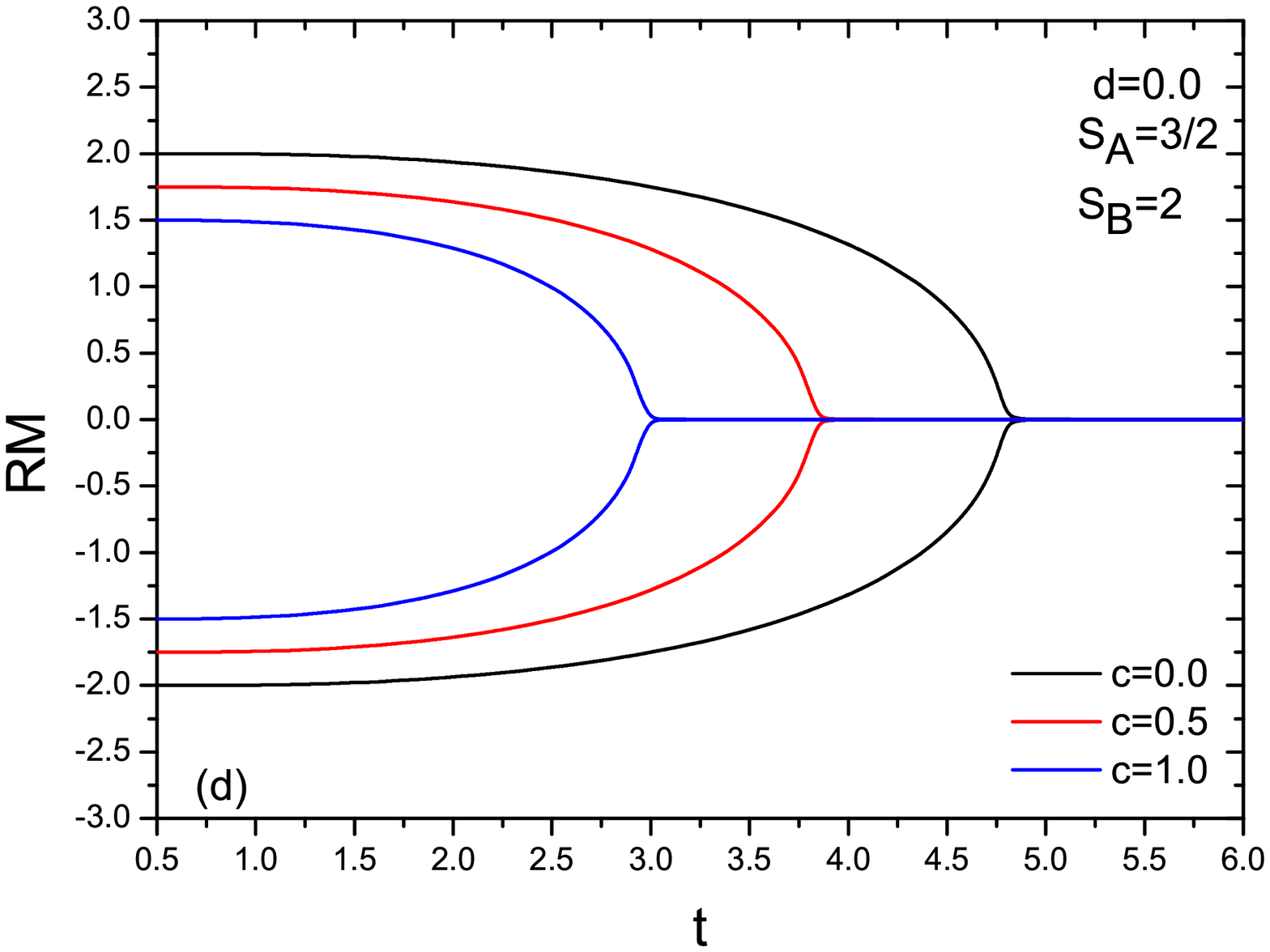, width=8cm}
\caption{Variation of RM with the temperature of the binary alloy system which consists of 
(a) $S_{A}=1/2$, $S_{B}=3/2$, (b) $S_{A}=1$, $S_{B}=2$, (c) $S_{A}=1$, $S_{B}=3/2$
and (d) $S_{A}=3/2$, $S_{B}=2$ spin variables for selected values of concentrations $c=0.0$, $c=0.5$ and $c=1.0$ at 
$d=0$ crystal field parameter.}\label{sek8}
\end{figure}

\begin{figure}[h]
\epsfig{file=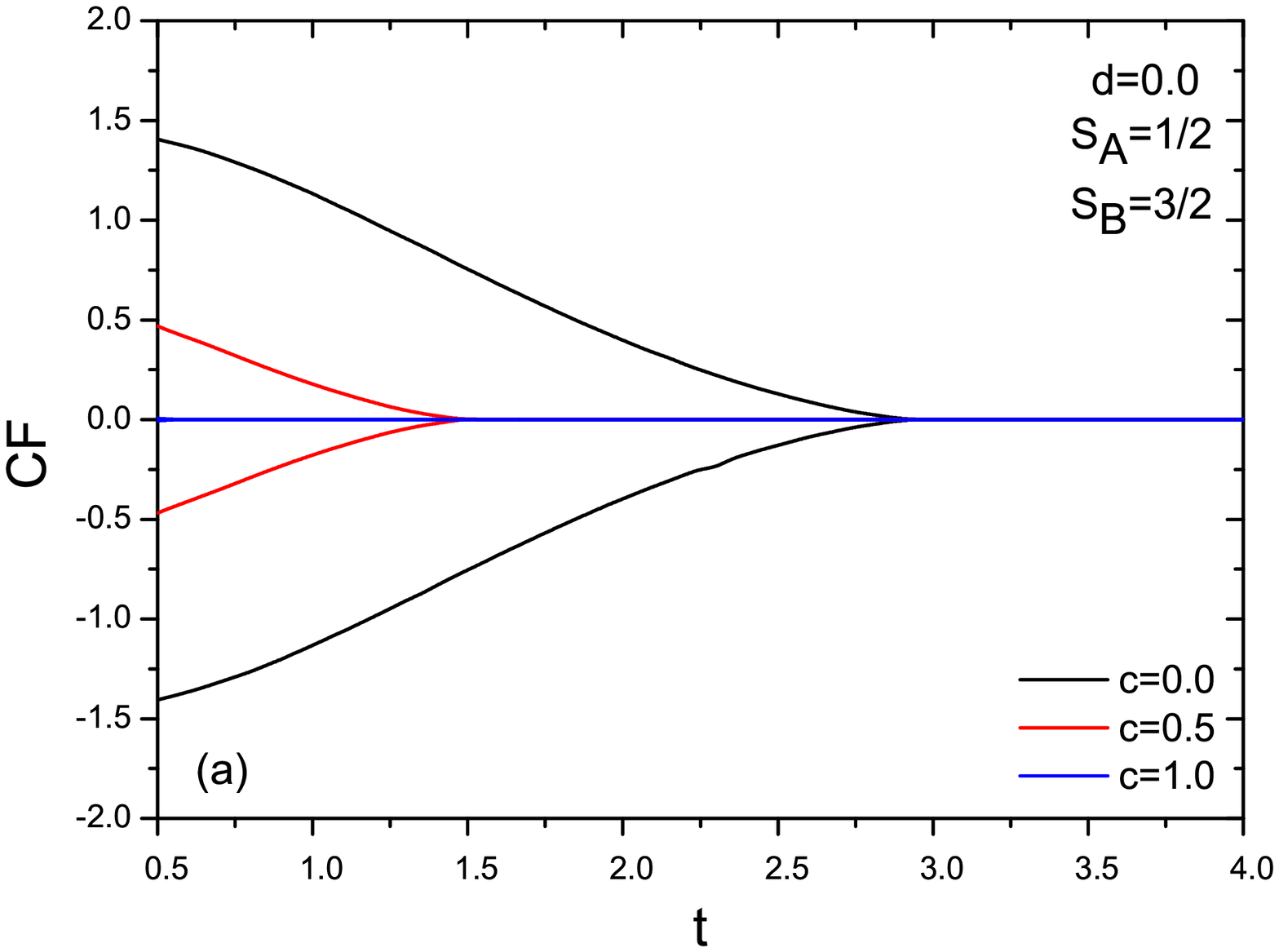, width=8cm}
\epsfig{file=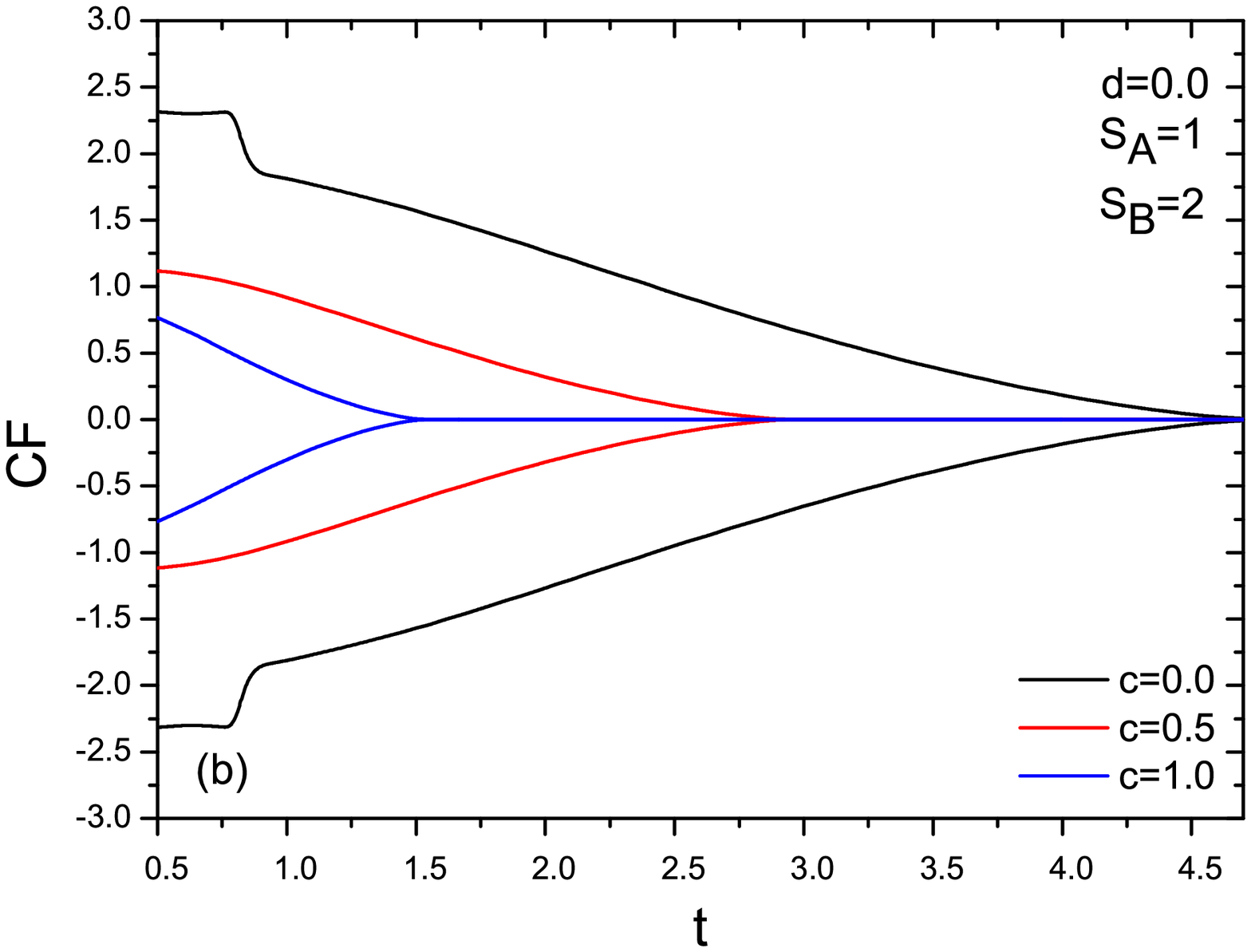, width=8cm}
\epsfig{file=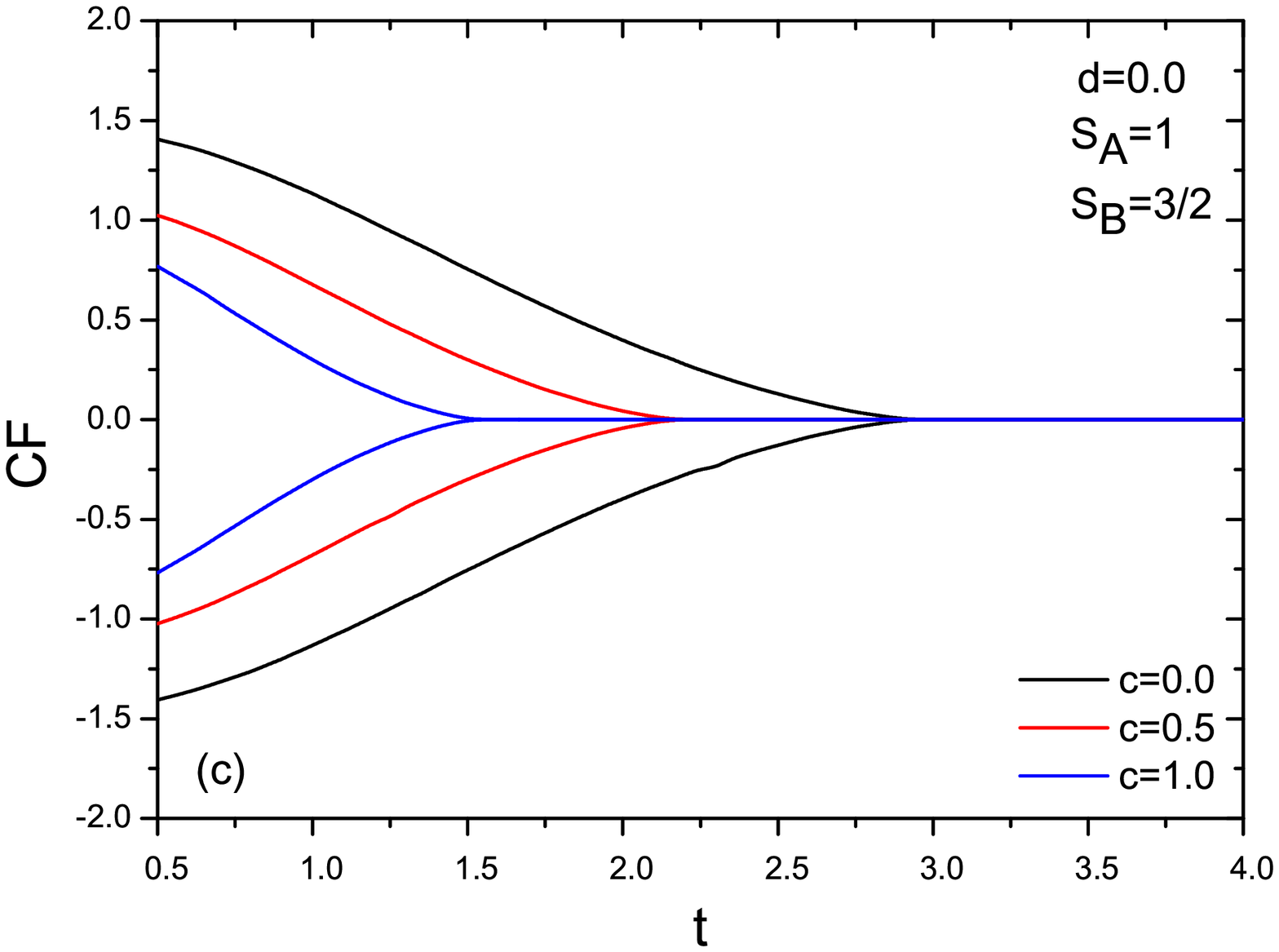, width=8cm}
\epsfig{file=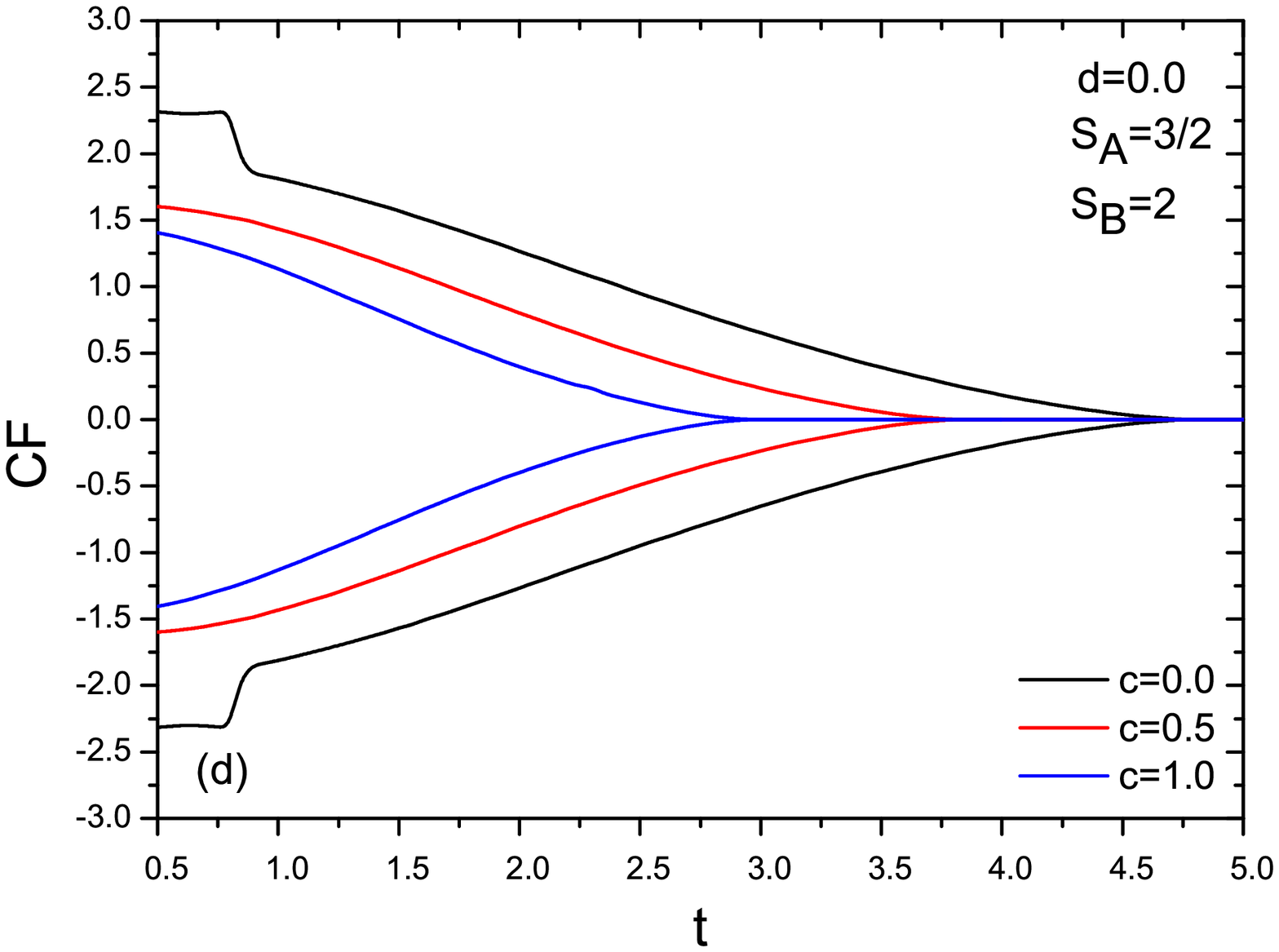, width=8cm}
\caption{Variation of CF with the temperature of the binary alloy system which consists of 
(a) $S_{A}=1/2$, $S_{B}=3/2$, (b) $S_{A}=1$, $S_{B}=2$, (c) $S_{A}=1$, $S_{B}=3/2$
and (d) $S_{A}=3/2$, $S_{B}=2$ spin variables for selected values of concentrations $c=0.0$, $c=0.5$ and $c=1.0$ at 
$d=0$ crystal field parameter.}\label{sek9}
\end{figure}

In Figs. \ref{sek7}, \ref{sek8} and  \ref{sek9} we demonstrate the variation of the quantities (HLA,
RM and CF, respectively) with the temperature, for one or both of two spin variables chosen as integer or half integer spin. Chosen spin values at crystal field $d=0$ are (a) $S_{A}=1/2$, $S_{B}=3/2$, (b) $S_{A}=1$, $S_{B}=2$, (c) $S_{A}=1$, $S_{B}=3/2$
and (d) $S_{A}=3/2$, $S_{B}=2$. Besides, the selected values of the concentrations are $c=0.0$, $c=0.5$ and $c=1.0$ in these figures. In Fig. \ref{sek7} (a) HLA drops
to zero as the temperature increases. Note that, the HLA of $c=0.0$ is greater than the HLA for $c=0.5$. In $c=1.0$ case there is no HLA, 
which means a paramagnetic hysteresis loop. Rising temperature drives the system to a disordered phase, due to the thermal
agitations. In Fig. \ref{sek7} (b) for integer-integer spin model, HLA exists for $c=1$ concentration at low temperatures and the value of HLA is greater than Fig. \ref{sek7} (a).
Therefore, more energy dissipation occurs. The temperature values that HLA vanishes, increases. For $c=0$ the value
of HLA has remained constant at low temperatures as in Fig \ref{sek7} (b). If we fix spin value of A atoms as $S_{A}=1$, and increase the 
spin value of B atoms such that $S_{B}=3/2$, HLA curves of $c=0.0$ and $c=0.5$ decrease (compare Fig \ref{sek7} (b) and (c)). 
If we fixed the spin value of B atoms as $S_{B}=2$, then increase the spin value of A atoms such that $S_{A}=3/2$, HLA curves of $c=0.5$ and $c=1.0$ 
get higher (compare Fig. \ref{sek7} (b) and (d)). As seen in Figs. \ref{sek8} for RM, and  Fig. \ref{sek9} for CF, decreasing behavior occurs with the increasing concentration. According to
the different spin variables, results of RM and CF are consistent with HLA's behavior that is obtained above.



\section{Conclusion}\label{conclusion}
In conclusion, hysteresis characteristics of the generalized spin-S magnetic binary alloy system represented
by $A_c B_{1-c}$ have been investigated within the framework of effective field theory. The system consists 
of type-$A$ and type-$B$ atoms with the concentrations $c$ and $1-c$, respectively.  Results of the generalized spin-$S$ binary alloy model are 
discussed as one or both of the spins of $A$ and $B$ atoms are selected as integer or half integer spin values.

The effects of the concentration and crystal field parameters of the magnetic binary alloy model strongly
depend on whether the spins of the atoms are integer or half-integer. As consistently by the related literature, the special 
cases ($c=0$, $c=1$) of binary alloy system exhibit $2S$-windowed
hysteresis character. The evolution of the multiple hysteresis loops are observed for higher spin valued alloy system for the 
large negative value of crystal field at low temperatures. Integer (half-integer) spin valued binary alloy 
system exhibits disordered (ordered) phase in this region. It has been found that
the number of outer windows that disappeared will be twice that difference between spin values
when majority of binary alloy composed of type A atoms in the case of $S_A<S_B$.

The significantly remarkable point of our results is that the effect of concentration is one of the most important
parameters affecting the hysteresis behavior in the system. Since the majority of the lattice sites consist
of half-integer spin values, the center loop lost and the system exhibits phase transition as the concentration
increases for integer-half integer system. Besides, the magnetic ground state of the system disappears in the low
magnetic field. As the applied magnetic field increases, the transition between the other ground states of 
the system occurs with a very small magnetic field value. We have decided that the difference of the magnetization values between 
the new ground states dominated by the B atom is less than $1$. For alloys with
concentration consisting mostly integer spin valued atoms, the ground state will be dominated by type-A atoms.
It is remarkable generalized result that $2S_{B}-2S_{A}-PH-2S_{A}$-windowed hysteresis loops are observed 
with increasing concentrations for integer-half integer system. The transition from $2S_{B}$ to 
$2S_{A}$-windowed loop causes $2(S_{B}-S_{A})$ vanishing windows. When the majority of binary alloy is composed 
of integer spin values, inner (low magnetic field) ground states of the system disappear at low magnetic field as  we started to
add more half-integer spin to the half integer-integer system. As concentration rises, the magnetic ground
states dominated by half-integer spin appear and the system is in an ordered phase anymore. 

It has been demonstrated that the outermost symmetric loops disappear gradually and windows are separated
from each other, as the crystal field parameter is increased in the negative direction. 
Besides, the quantities of hysteresis loops have been investigated
with the variation of the temperature. Rising temperature drags the system into a disordered phase due to the  
thermal agitations. As the concentration increases from $c=0$ to $1$, HLA, RM and CF
decreases for all binary alloy system which is one or both of two spin variables chosen as integer or half 
integer spin model. These quantities increase as the spin value gets higher. 

We hope that the results obtained in this work may be beneficial form both theoretical and experimental 
points of view.





\bibliographystyle{model1-num-names}
\bibliography{<your-bib-database>}

\end{document}